\documentclass[%
 reprint,
superscriptaddress,
nofootinbib,
 amsmath,amssymb,
 aps,onecolumn
]{revtex4-2}
\usepackage{color}
\usepackage{stmaryrd}
\usepackage{graphicx}
\usepackage{dcolumn}
\usepackage{bm}
\usepackage{physics} 

\newcommand{\tj}[6]{ \begin{pmatrix}
  #1 & #2 & #3 \\
  #4 & #5 & #6 
 \end{pmatrix}}

\newcommand{\dstrike}{\partial_{r_\star}\!\!\!\!\!\!\!\backslash\,~~}

\begin{document}

\title{Corrections to Hawking Radiation from Asteroid Mass Primordial Black Holes: \\
I. Formalism of Dissipative Interactions in Quantum Electrodynamics} 


\author{Makana Silva}
\email{silva.179@osu.edu}
\author{Gabriel Vasquez}%
\email{vasquez.119@osu.edu}
\author{Emily Koivu}
\email{koivu.1@osu.edu}
\author{Arijit Das}
\email{das.241@osu.edu}
\affiliation{%
 Center for Cosmology and Astroparticle Physics, The Ohio State University,
 191 West Woodruff Avenue, Columbus OH, 43210, USA
}
\affiliation{%
 Department of Physics, The Ohio State University,
 191 West Woodruff Avenue, Columbus OH, 43210, USA
}
\author{Christopher M. Hirata}
 \email{hirata.10@osu.edu}
\affiliation{%
 Center for Cosmology and Astroparticle Physics, The Ohio State University,
 191 West Woodruff Avenue, Columbus OH, 43210, USA
}
\affiliation{%
 Department of Physics, The Ohio State University,
 191 West Woodruff Avenue, Columbus OH, 43210, USA
}
\affiliation{%
 Department of Astronomy, The Ohio State University,
 140 West 18th Avenue, Columbus OH, 43210, USA
}

\date{October 4, 2022}

\begin{abstract}
Primordial black holes (PBHs) within the mass range $10^{17} - 10^{22}$ g are a favorable candidate for describing the all of the dark matter content. Towards the lower end of this mass range, the Hawking temperature, $T_{\rm H}$, of these PBHs is $T_{\rm H} \gtrsim 100$ keV, allowing for the creation of electron -- positron pairs; thus making their Hawking radiation a useful constraint for most current and future MeV surveys. This motivates the need for realistic and rigorous accounts of the distribution and dynamics of emitted particles from Hawking radiation in order to properly model detected signals from high energy observations. This is the first in a series of papers to account for the $\mathcal{O}(\alpha)$ correction to the Hawking radiation spectrum. We begin by the usual canonical quantization of the photon and spinor (electron/positron) fields on the Schwarzschild geometry. Then we compute the correction to the rate of emission by standard time dependent perturbation theory from the interaction Hamiltonian. We conclude with the analytic expression for the \textit{dissipative} correction, i.e. corrections due to the creation and annihilation of electron/positrons in the plasma. 
\end{abstract}

\maketitle

\section{Introduction}

Primoridal black holes (PBHs) are possible relics that could provide insights into the physics of the earliest moments of the Universe \cite{1966AZh....43..758Z, 1971MNRAS.152...75H, 1974MNRAS.168..399C, 1975ApJ...201....1C, 1975Natur.253..251C}. There are several proposed formation mechanisms for PBHs, such as collapse of non-Gaussian fluctuations in the Early Universe, tunneling through some scalar potential, etc. \cite{1971MNRAS.152...75H, 2010JCAP...04..023F, 2016PhRvD..94h3504C, 2019JCAP...10..077C, 2022arXiv220407152L, 2022PhRvD.106b3519P}. Primordial black holes are an interesting candidate for dark matter (DM) since, although some new physics at a high energy scale is required to form them, the PBH scenario does not require any new long-lived particles to be added to the Standard Model \cite{2022arXiv220908215D}. A wide range of observational constraints have removed different mass ranges as candidates for dark matter (see recent summaries \cite{2019JCAP...08..031M,2021RPPh...84k6902C}), leaving a mass range of about $10^{17}$--$10^{22}$ g as a possible candidate to describe all of the dark matter content. 

There are several probes for detecting PBHs based on their associated physical properties. From a gravitational perspective, PBHs could be detected through their lensing effects on various bright back ground sources, e.g. microlensing, gravitational wave detection, etc. \cite{2016PhRvL.116t1301B, 2019RNAAS...3...58L, 2019JCAP...08..031M, 2020JCAP...09..022J, 2017PDU....18..105C, 2017PhRvD..95d3511N, 2018PhR...761....1B, 2018PhLB..782...77T}. Another method involves quantum processes near the black hole horizon that were predicted by \citet{1975CMaPh..43..199H} known as \textit{Hawking radiation}. This is the predicted effect that black holes would radiate away (``evaporation'') by the emission of radiation and other particles. The foundation of this argument lies in how we define the vacuum state; prior to the formation of the black hole (flat spacetime) we define a vacuum state with zero occupation of particles, but after a black hole forms, the vacuum near the horizon becomes a thermal state (due to the existence of an accelerating frame of reference near the horizon, i.e. Unruh effect) from which particles can arise and escape to infinity \cite{1976PhRvD..14..870U,1975CMaPh..43..199H, 1977RSPSA.356..237D, 1982PhRvD..25..942U}.  The Hawking radiation of a black hole is dependent on its mass: $T_H = 1/(8 \pi M)$, where $T_H$ is the Hawking temperature of the radiation in units where $G = k_B = c = \hbar = 1$, showing that lower mass black holes radiate at higher temperatures than more massive ones. This process places constraints on the lower mass range of PBHs due to lifetime of PBHs with $M_{PBH} \lesssim 5 \times 10^{14}$ g being comparable to the age of the Universe, i.e. evaporated away \cite{1975ApJ...201....1C, 2022arXiv220305743M}. For PBHs of $M_{PBH} \lesssim 10^{17}$ g, their Hawking radiation would be in the $\gamma$-ray regime, making it a novel and effective probe for direct detection \cite{2021PhRvL.126q1101C, 2022arXiv220602672A}. 

This particular mass range (``asteroid mass'') and lower would have a Hawking temperature $T_H \gtrsim 100$ keV ($T_H \sim (10^{16}~{\rm g}/M_{PBH})$ MeV), well within the $\gamma$-ray regime of the electromagnetic spectrum, making their Hawking radiation a point of interest for observations by MeV observatories such as AMEGO or SMILE \cite{2020SPIE11444E..67T, 2019BAAS...51g.245M, 2021PhRvD.104b3516R, 2022PhRvD.106d3003K}. For the higher regime of $T_{\rm H}$ (i.e. lower end of the asteroid mass regime), PBHs are able to emit electron -- positron ($e^+e^-$) pairs, making them the dominate contributor to the detectable Hawking radiation and allowing for constraints on PBH mass distributions. Since the emission of $e^+e^-$ would lead to the increase of flux of 511 keV lines in any MeV survey and PBHs occupy spherical halos around galaxies, the 511 keV emission line near and away from the the Galactic center would be an ideal method of constraint \cite{2021PhRvD.104f3033K}. \citet{2019PhRvL.123y1102D} was able to place constraints on PBH abundance as a DM candidate by using 511 keV lines from MeV surveys of the Galactic Bulge from INTEGRAL \cite{2016A&A...586A..84S}, though the underlying assumption was the rate of production of $e^+e^-$ was not enough to form a plasma around a PBH (similar to \cite{2008PhRvD..78f4043M}), thus allowing for the escape of positrons to be annihilated at some further distance. This extended travel of the positron requires knowledge of its propagation through the ISM in order to carefully account for excess 511 keV lines in various regions of the Galactic Bulge. Recently, \citet{2021PhRvL.126q1101C} showed that upcoming MeV telescopes could directly detect asteroid mass PBHs within dark matter halos. The dependence of the particles created by PBHs at $T_{\rm H} \gtrsim 100$ keV motivates the need to have careful and rigorous models of the distributions of particles in order to use intermediate-energy leptons \cite{2019PhRvL.122d1104B} and the 511 keV positron annihilation line \cite{2019PhRvL.123y1102D, 2019PhRvL.123y1101L, 2020PhRvD.101l3514L} to constrain PBH models of DM.

Although Hawking radiation is often described as a blackbody, the emission spectrum is modified by a ``graybody'' factor related to the energy-dependent cross section for the black hole to absorb a particle. The standard calculation treats this by the partial wave expansion for non-interacting particles \cite{1976PhRvD..13..198P, 1976PhRvD..14.3260P}. Since then, there have been many investigations into the case of interactions and secondary particles in various approximations. \citet{1977PhRvD..16.2402P} showed that when charged particles are emitted, the black hole develops an opposite charge, leading to an electric potential that alters the emission of further charged particles, and an ${\cal O}(\alpha)$ change to the charged particle emission rate, where $\alpha \approx 1/137$ is the fine structure constant. At $T_{\rm H}\gtrsim 20\,$MeV, strongly interacting particles can be produced, and there have been investigations of hadronization \cite{1990PhRvD..41.3052M, 1991PhRvD..44..376M}. Also most of the particles produced at these higher energies are unstable, leading to secondary particles from their decay; in some regimes, these processes can even dominate the neutrino spectrum \cite{1995PhRvD..52.3239H}. Several papers have discussed the possibility that at very high $T_{\rm H}$, the density of emitted particles might result in a high optical depth and form a ``photosphere'' or a region of relativistic fluid flow followed by decoupling \cite{1997PhRvD..55..480H, 1997PhRvL..78.3430H, 1999PhRvD..59f3009C, 2001PhRvL..86.1670K, 2002PhRvD..65f4028D}, although subsequent work incorporating the special relativistic kinematics of the outgoing particles showed a much smaller optical depth and no photosphere \cite{2008PhRvD..78f4043M}. \citet{2008PhRvD..78f4044P} considered the ${\cal O}(\alpha)$ inner bremsstrahlung emission accompanying the emission of charged particles and showed that this process could be a significant contribution to the low-energy spectrum.
\citet{2021PhRvL.126q1101C} specifically show that the secondary spectrum (lower energy photons) of Hawking radiation from PBHs in the MeV range would be the most significant contributor to the detection of this spectrum of radiation, showing that the discovery reach for upcoming MeV surveys will increase by an order of magnitude by including the secondary spectrum. This shows that understanding the secondary spectrum is vital for any sort of MeV survey that hopes to make direct detection of Hawking radiation from PBHs. This has also motivated study of other quantum electrodynamics (QED) contributions to the emitted spectrum such as annihilation of $e^+e^-$ pairs \cite{2022PTEP.2022c3E03S}.

This is the first in a series of papers whose ultimate goal is to compute the ${\cal O}(\alpha)$ correction to the Hawking radiation from a Schwarzschild black hole. The current analysis of the photon spectrum from PBHs was investigated to ${\cal O}(\alpha)$ in \citet{2021PhRvL.126q1101C} showing at lower photon energies, the secondary spectrum becomes the dominant contributor to the total spectrum (see Fig.~2 in \citet{2021PhRvL.126q1101C}). However, this photon emission calculation was performed using flat spacetime inner bremsstrahlung formulae \cite{2020JCAP...01..056C}, leading to the purpose and motivation of this series of papers: to compute the ${\cal O}(\alpha)$ correction to the Hawking radiation but on a curved spacetime, i.e., the Schwarzschild metric. Given the historical challenges and controversies in understanding interacting particles resulting from Hawking radiation, we believe that a full quantum treatment on the curved background is a necessary step in order to put the calculation of inner bremsstrahlung and related $\mathcal{O}(\alpha)$ effects on a rigorous foundation.

This paper is organized as follows. In Sec.~\ref{sec: Conventions} we lay out our conventions for the spacetime metric and spinor algebra used in the quantization of the electromagnetic and spinor fields (electron and positrons). In Sec.~\ref{sec: EM_field} and Sec.~\ref{sec: electron_field}, we work through the canonical quantization of the electromagnetic and spinor fields. In Sec.~\ref{sec: H_int}, we compute the interaction Hamiltonian between spinors and photons in terms of annihilation and creation operators and mode overlap integrals. In Sec.~\ref{sec: photon_evolve}, we re-express our results in the interaction picture as appropriate for a time dependent perturbation theory treatment, and arrive at our expressions for the dissipative ${\cal O}(\alpha)$ contribution to the photon emission of the Hawking radiation. Finally, we conclude and discuss follow up works in Sec.~\ref{sec: Discussion}.

\section{Conventions}
\label{sec: Conventions}

We use the $+++-$ signature for the metric. We use Greek indices $\alpha\beta$ to indicate any spacetime index; and Latin indices $ijk$ to select only spatial indices; and Latin indices $ABC$ to denote Dirac spinor indices.

We use the tortoise coordinate $r_\star$ to write the Schwarzschild metric:
\begin{equation}
ds^2 = \left( 1 - \frac{2M}r \right) dr_\star^2 + r^2\,d\theta^2 + r^2\sin^2\theta\,d\phi^2 -\left( 1 - \frac{2M}r \right)dt^2,
\label{eq:Schwarzschild}
\end{equation}
where $r$ and $r_\star$ are related by
\begin{equation}
r_\star = r + 2M \ln \frac{r-2M}{2M}
~~~{\rm and}~~~
\frac{dr}{dr_\star} = 1-\frac{2M}r.
\end{equation}
We have $r_\star \approx r$ at large radius, but at the horizon $r\rightarrow 2M$ whereas $r_\star \rightarrow -\infty$. In general $r$ appears in most of our expressions, but $r_\star$ is a better choice of independent variable for numerical calculations near the horizon since one can avoid cancellations in the expression $1-2M/r$.

We use the overdot symbol $\,\dot{}\,$ to indicate partial derivatives with respect to $t$, and the prime ${}'$ to indicate partial derivatives with respect to $r_\star$.

The electron mass is $\mu$ and the black hole mass is $M$. The speed of light, Newton's gravitational constant, the reduced Planck's constant (``$\hbar$''), and the permittivity of the vacuum (``$\epsilon_0$'') are set equal to unity.

The Dirac matrices written in an orthonormal basis are denoted by $\tilde{\boldsymbol\gamma}_\mu$, and satisfy the usual anti-commutation relation $\{\tilde{\boldsymbol\gamma}_\mu,\tilde{\boldsymbol\gamma}_\nu\} = 2\eta_{\mu\nu} \mathbb I_{4 \times 4}$. We use the representation of the Dirac matrices, in $2\times 2$ form,
\begin{equation}
\tilde{\boldsymbol\gamma}_i = \left( \begin{array}{cc} 0 & {\boldsymbol\sigma}_i \\ {\boldsymbol\sigma}_i & 0 \end{array}\right)
~~{\rm and}~~
\tilde{\boldsymbol\gamma}_4 = \left( \begin{array}{cc} i{\mathbb I}_{2\times 2} & 0 \\ 0 & -i{\mathbb I}_{2\times 2} \end{array}\right),
\label{eq: gamma_matrices}
\end{equation}
where ${\boldsymbol\sigma}_i$ are the Pauli matrices and ${\mathbb I}_{2\times 2}$ is the $2\times 2$ identity. This follows the convention of Brill \& Wheeler \cite{1957RvMP...29..465B}; note that due to the signature, $\tilde{\boldsymbol\gamma}_4$ is anti-Hermitian whereas $\tilde{\boldsymbol\gamma}_i$ is Hermitian. We define the usual adjoint spinor appearing in the Dirac Lagrangian $\bar\psi = \psi^\dagger{\boldsymbol\beta}$, where 
\begin{equation}
{\boldsymbol\beta}=-i\tilde{\boldsymbol\gamma}_4
= \left( \begin{array}{cc} {\mathbb I}_{2\times 2} & 0 \\ 0 & -{\mathbb I}_{2\times 2} \end{array}\right).
\label{eq:betadef}
\end{equation}

In this convention, the action of quantum electrodynamics is
\begin{equation}
S_{\rm QED} = \int \left[
 \bar\psi\left(-{\boldsymbol\gamma}^\mu D_\mu - \mu \right)\psi
-\frac14F_{\mu\nu}F^{\mu\nu} \right] \sqrt{-g}\,d^4x,
\label{eq:action-QED}
\end{equation}
where ${\boldsymbol\gamma}^\mu = a^\mu{_\alpha}\tilde{\boldsymbol\gamma}^\alpha$ is the Dirac ${\boldsymbol\gamma}$-matrix in covariant notation; $a^\mu{_\alpha}$ is the $4\times 4$ matrix of vierbein components; and the covariant derivative acting on the electron spinor is
\begin{equation}
D_\mu = \partial_\mu - {\boldsymbol\Gamma}_\mu - ieA_\mu.
\label{eq:covariant-derivative}
\end{equation}
This expression contains two corrections to the partial derivative: the $4\times 4$ ${\boldsymbol\Gamma}_\mu$ matrix, which encodes the rotation of the vierbein when we move in direction $\mu$ (Eq.~8 of Ref.~\cite{1957RvMP...29..465B}); and the $U(1)$ gauge transformation term $-ieA_\mu$ for electron charge $e$. The electromagnetic term includes the field strength tensor, $F_{\mu\nu} = \partial_\mu A_\nu - \partial_\nu A_\mu$.

We denote quantum numbers of modes as follows:
\begin{enumerate}
    \item For {\em photon modes}, the energy is denoted $\omega$ and the angular quantum numbers are denoted by $\ell$ and $m$. Parity is denoted by $(p)\in \{(e),(o)\}$ (``even'' or ``odd'' sector; also described as ``electric'' or ``magnetic'' when discussing atomic or nuclear transitions, or ``polar'' or ``axial'' sectors \cite{1998mtbh.book.....C}).
    \item For {\em electron modes}, the energy is denoted by $h$ and the total (orbital+spin) angular momentum and parity are denoted by $j$, $m$, and $p=\pm 1$. We may for shorthand use the combination $k = s(j+\frac12)$ instead of writing both $j$ and $p$, where the sign of $k$ is $s = (-1)^{j-1/2}p$.
    \item For both types of modes, we use $X$ to denote a mode associated with the ``in from $\infty$'' (``in'') channel, or ``up from the horizon'' (``up'') channel, as depicted in Fig.~\ref{fig: modes}. We may also use the alternative basis of the ``out'' or ``down'' modes. 
    \begin{figure}
    \centering
    \includegraphics[scale=0.2]{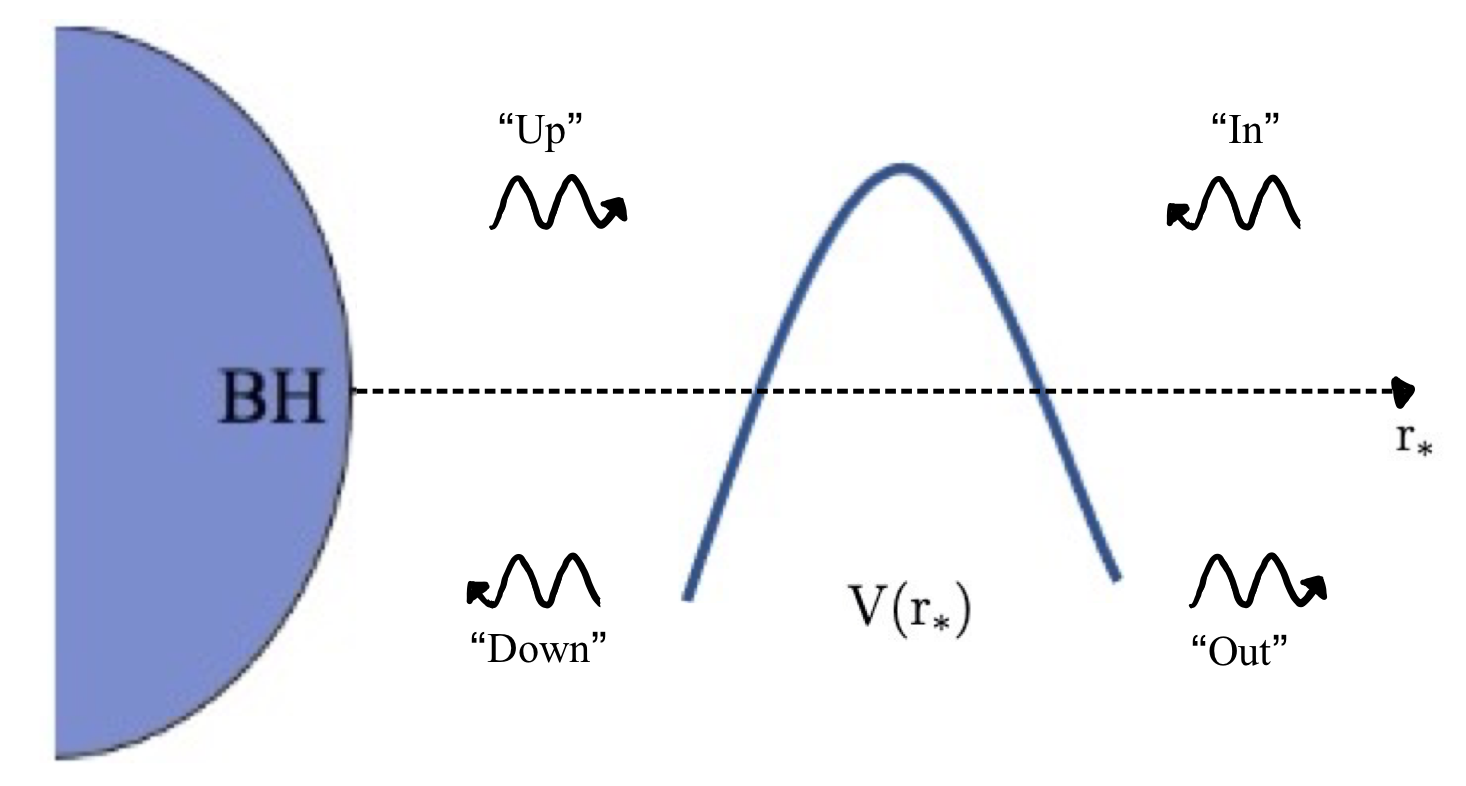}
    \caption{Representation of the different scattering states from the effective potential (curve labeled $V(r_\ast)$) coming from the black  hole (BH) horizon (``up'') and coming in from $\infty$ (``in'') as well as the equivalent scattering basis for radiation falling towards the BH near the horizon (``down'') and radiation leaving the black hole to $\infty$ (``out'').}
    \label{fig: modes}
\end{figure}
    \item Since we will be doing non-linear interactions, we will often work with multiple electron or photon modes. Mode indices may be primed or subscripted to distinguish them.
\end{enumerate}

We define the triangle inequality condition:
\begin{equation}
\Delta(j_1,j_2,j_3) 
= \left\{\begin{array}{ccc}
1 & & j_1+j_2\ge j_3 {\rm ~~and~~} j_3+j_1\ge j_2
{\rm ~~and~~} j_2+j_3\ge j_1 \\
0 & & {\rm otherwise}
\end{array}\right.,
\label{eq:Triangle}
\end{equation}
and define $(-1)^p = +1$ for $(p)=(e)$ and $(-1)^p=-1$ for $(p)=(o)$.

\section{The electromagnetic field}
\label{sec: EM_field}

We first review the electromagnetic field in the Schwarzschild spacetime. This is described by an Schr\"odinger-like equation with a Regge-Wheeler potential \cite{1955PhRv...97..511W, 1957PhRv..108.1063R}, which is implemented in commonly used tools such as {\tt BlackHawk} to compute Hawking radiation for spin 1 massless particles (see Eq.~3.33 of \cite{2021PhRvD.103j4010A}). We work through the derivation here, both for completeness and to make explicit our gauge choice and the expression for potentials in terms of annihilation and creation operators.

Because we want to write an interaction Hamiltonian, we will follow the canonical quantization method, and we also work in terms of the vector potential $A_\mu$ rather than the field components $F_{\mu\nu}$ (or the closely related Newman-Penrose variables $\Phi_0$ and $\Phi_2$ \cite{1962JMP.....3..566N, 1972PhRvL..29.1114T, 1973ApJ...185..635T, 1974ApJ...193..443T}). This approach has been used for some problems in black hole spacetimes, although with a different gauge choice than we will take here -- e.g., the modified Feynman gauge \cite{2001PhRvD..63l4008C}, the gauge choices based on null vectors \cite{1975PhRvD..11.2042C, 1981PhRvD..24..297C} (which extend naturally to the Kerr case), or the zero scalar potential gauge $A_t=0$ \cite{1998PhRvD..57.1108C} (although this leads to the same mode functions for the vector potential because it is the same as the Coulomb gauge in the case where there are no charges). It is also used for the Proca equation \cite{2012PhRvD..85b4005H, 2012PhRvD..85d4043R} (but in that case the Lorenz condition $A^\mu{_{;\mu}}=0$ holds and we cannot impose a different gauge choice).

\subsection{Angular decomposition}

The electromagnetic field action is $S_{\rm EM} = \int -\frac14 F_{\mu\nu}F^{\mu\nu}\,\sqrt{-g}\,d^4x$, where the field tensor is $F_{\mu\nu} = \partial_\mu A_\nu - \partial_\nu A_\mu$. Written in terms of components, the Lagrangian $L_{\rm EM} = dS_{\rm EM}/dt$ is
\begin{eqnarray}
L_{\rm EM} &=& \int_{-\infty}^\infty \int_0^{2\pi}\int_0^\pi \Bigl\{ \frac12(\Phi' + \dot A_{r_\star})^2
+ \frac{1-2M/r}{2r^2} \Bigl[ (\partial_\theta\Phi + \dot A_{\theta})^2 +\frac1{\sin^2\theta} (\partial_\phi\Phi + \dot A_{\phi})^2 \Bigr]
\nonumber \\ &&
- \frac{1-2M/r}{2r^2} \Bigl[ (\partial_\theta A_{r_\star} - A'_\theta)^2 + \frac1{\sin^2\theta} (\partial_\phi A_{r_\star} - A'_\phi)^2 \Bigr]
- \frac{(1-2M/r)^2}{2r^4\sin^2\theta} (\partial_\theta A_{\phi} - \partial_\phi A_\theta)^2
\Bigr\}\,\sin\theta\,d\theta\,d\phi\,
\frac{r^2 \,dr_\star}{1-2M/r},
\nonumber \\ &&
\label{eq:LEM}
\end{eqnarray}
where $\Phi \equiv -A_t$. It is convenient to break down the vector potential in a multipole expansion: we write
\begin{eqnarray}
\Phi(t,r_\star,\theta,\phi) &=& \sum_{\ell=0}^\infty \sum_{m=-\ell}^\ell \Phi_{\ell,m}(t,r_\star) Y_{\ell,m}(\theta,\phi),
\nonumber \\
A_{r_\star}(t,r_\star,\theta,\phi) &=& \sum_{\ell=0}^\infty \sum_{m=-\ell}^\ell A_{(r),\ell,m}(t,r_\star) Y_{\ell,m}(\theta,\phi),
\nonumber \\
A_{\theta}(t,r_\star,\theta,\phi) &=& \sum_{\ell=1}^\infty \sum_{m=-\ell}^\ell \left[ A_{(e),\ell,m}(t,r_\star) \partial_\theta Y_{\ell,m}(\theta,\phi)
- A_{(o),\ell,m}(t,r_\star) \frac1{\sin\theta} \partial_\phi Y_{\ell,m}(\theta,\phi) \right],
~~~{\rm and}
\nonumber \\
A_{\phi}(t,r_\star,\theta,\phi) &=& \sum_{\ell=1}^\infty \sum_{m=-\ell}^\ell \left[ A_{(e),\ell,m}(t,r_\star) \partial_\phi Y_{\ell,m}(\theta,\phi)
+ A_{(o),\ell,m}(t,r_\star) \sin\theta\, \partial_\theta Y_{\ell,m}(\theta,\phi) \right].
\label{eq:A-Multipole}
\end{eqnarray}
Here $\Phi_{\ell,m}$ and $A_{(r),\ell,m}$ are the usual multipole moments of the scalar potential and radial part of the vector potential. The components in the angular ($\theta$ and $\phi$) directions form a vector on the 2-sphere of constant $(t,r_\star)$ and have to be described with two sets of multipole moments: an $A_{(e),\ell,m}$ component with the same parity as $Y_{\ell,m}$ (since the coefficient multiplies $\nabla Y_{\ell,m}$); and an $A_{(o),\ell,m}$ component with the opposite parity (since the coefficient multiplies $\hat{\bf e}_{r_\star}\times\nabla Y_{\ell,m}$). An equivalent description of the angular components is
\begin{equation}
A_{\theta}(t,r_\star,\theta,\phi) \pm \frac{i}{\sin\theta} A_\phi(t,r_\star,\theta,\phi) = \mp \sum_{\ell=1}^\infty \sum_{m=-\ell}^\ell
\sqrt{\ell(\ell+1)}\,\left[
A_{(e),\ell,m}(t,r_\star) \pm iA_{(o),\ell,m}(t,r_\star) \right]\, {_{\pm 1}}Y_{\ell,m}(\theta,\phi).
\end{equation}
The Lagrangian of Eq.~(\ref{eq:LEM}) can be re-written in terms of these components; with some simplification, we find
\begin{eqnarray}
L_{\rm EM} &=& \sum_{\ell,m} \int_{-\infty}^\infty \Bigl\{ \frac{r^2}{1-2M/r}
\Bigl[\frac12|\Phi'_{\ell,m}|^2 + \Re {\Phi'}^\ast_{\ell,m} \dot A_{(r),\ell,m} + \frac12 |\dot A_{(r),\ell,m}|^2\Bigr]
+ \ell(\ell+1)\Bigl[ \frac12 |\Phi_{\ell,m}|^2+ \Re \Phi^\ast_{\ell,m} \dot A_{(e),\ell,m}
\nonumber \\ && \!\!
 + \frac12 |\dot A_{(e),\ell,m}|^2
+ \frac12 |\dot A_{(o),\ell,m}|^2
-\frac12 |A_{(r),\ell,m} - A'_{(e),\ell,m}|^2
- \frac12 |A'_{(o),\ell,m}|^2
 \Bigr]
- [\ell(\ell+1)]^2 \frac{1-2M/r}{r^2} \frac12 |A_{(o),\ell,m}|^2
\Bigr\}dr_\star.
\nonumber \\ &&
\label{eq:LEM2}
\end{eqnarray}

\subsection{Gauge fixing}

To continue the quantization program, we impose a gauge condition. A particularly convenient choice is a generalization of the Coulomb gauge, which eliminates the cross-terms between the scalar and vector potentials. In Minkowski spacetime the Coulomb gauge has the disadvantage of breaking Lorentz invariance, but on a Schwarzschild background this is not a concern. In our case, we choose
\begin{equation}
K_{\ell m} \equiv -\left(\frac{r^2}{1-2M/r} A_{(r),\ell,m}\right)' + \ell(\ell+1) A_{(e),\ell,m} = 0,
\label{eq:Coulomb}
\end{equation}
which turns the combination of the $\Re {\Phi'}^\ast_{\ell,m} \dot A_{(r),\ell,m} $ and $\Re \Phi^\ast_{\ell,m} \dot A_{(e),\ell,m}$ terms in Eq.~(\ref{eq:LEM2}) into a total derivative. The gauge choice of Eq.~(\ref{eq:Coulomb}) can always be achieved by the gauge transformation $A_\mu\rightarrow A_\mu + \partial_\mu\chi$, where
$\chi =- \tilde{\cal H}_\ell^{-1} K_{\ell m} Y_{\ell,m}(\theta,\phi)$
and we define the radial operator
\begin{equation}
\tilde {\cal H}_\ell f = -\left(\frac{r^2}{1-2M/r}f'\right)' + \ell(\ell+1)f,
\label{eq:HL-tilde}
\end{equation}
which is positive definite and Hermitian with respect to the standard inner product $(f_1|f_2)_{r_\star} = \int_{-\infty}^\infty f_1^\ast f_2\,dr_\star$.

\subsection{Sectors of the theory and the radial wave equations}

The Lagrangian of Eq.~(\ref{eq:LEM2}) can then be broken down into three parts:
\begin{equation}
L_{\rm EM} = L_{{\rm EM},\Phi} + L_{{\rm EM},(e)} + L_{{\rm EM},(o)},
\end{equation}
where the scalar potential, even-parity vector potential, and odd-parity vector potential parts are
\begin{eqnarray}
L_{{\rm EM},\Phi} &=& \sum_{\ell,m} \int_{-\infty}^\infty \Bigl\{ \frac12\frac{r^2}{1-2M/r}|\Phi'_{\ell,m}|^2 + \frac12 \ell(\ell+1) |\Phi_{\ell,m}|^2\Bigr\}\,dr_\star,
\nonumber \\
L_{{\rm EM},(e)} &=& \sum_{\ell,m} \int_{-\infty}^\infty  \Bigl\{ \frac{r^2}{1-2M/r}
 \frac12 |\dot A_{(r),\ell,m}|^2
+ \ell(\ell+1)\Bigl[  \frac12 |\dot A_{(e),\ell,m}|^2
-\frac12 |A_{(r),\ell,m} - A'_{(e),\ell,m}|^2
 \Bigr]
\Bigr\}\,dr_\star,
~~{\rm and}
\nonumber \\
L_{{\rm EM},(o)} &=& \sum_{\ell,m} \ell(\ell+1) \int_{-\infty}^\infty \Bigl\{ 
 \frac12 |\dot A_{(o),\ell,m}|^2 -  \frac12 |A'_{(o),\ell,m}|^2 - \frac{\ell(\ell+1)(1-2M/r)}{2r^2} |A_{(o),\ell,m}|^2
\Bigr\}\,dr_\star.
\label{eq:LEM3}
\end{eqnarray}
The scalar potential part has no dynamics; when coupled to charges it will give rise to an instantaneous Coulomb-like interaction energy, just as happens in the Coulomb gauge in Minkowski space. The odd-parity vector potential part can be simplified by introducing the positive definite Hermitian operator
\begin{equation}
 {\cal H}_\ell f = -f'' + \frac{\ell(\ell+1)(1-2M/r)}{r^2}f,
\label{eq:HL}
\end{equation}
so that
\begin{equation}
L_{{\rm EM},(o)} = \sum_{\ell,m} \ell(\ell+1) \int_{-\infty}^\infty \Bigl\{ 
 \frac12 |\dot A_{(o),\ell,m}|^2 - \frac12 A^\ast_{(o),\ell,m}  {\cal H}_\ell A_{(o),\ell,m} \Bigr\}\,dr_\star.
\label{eq:Ao-1}
\end{equation}

It turns out that the even-parity potential part can be written in a similar form. This is not surprising, since we know that the electromagnetic field equations in vacuum do not distinguish electric vs.\ magnetic fields. We define the new variable $Z_{\ell,m}(t,r_\star)$ by
\begin{equation}
Z_{\ell,m} \equiv A_{(r),\ell,m}  - A'_{(e),\ell,m} 
=  A_{(r),\ell,m} -\frac1{\ell(\ell+1)}\left(\frac{r^2}{1-2M/r} A_{(r),\ell,m}\right)'',
\end{equation}
where the second equality comes from Eq.~(\ref{eq:Coulomb}). Thus $A_{(r),\ell,m}$ is related to $Z_{\ell,m}$ by a second-order differential equation; one can straightforwardly verify that the solution is
\begin{equation}
A_{(r),\ell,m} = \frac{\ell(\ell+1)(1-2M/r)}{r^2} {\cal H}_\ell^{-1}Z_{\ell,m},
\label{eq:Ar-from-Z}
\end{equation}
where ${\cal H}_\ell^{-1}Z_{\ell,m}$ denotes the inverse operation, i.e., the solution $f$ to ${\cal H}_\ell f = Z_{\ell,m}$. It follows that
\begin{equation}
A_{(e),\ell,m} = ({\cal H}_\ell^{-1}Z_{\ell,m})'.
\label{eq:Ae-from-Z}
\end{equation}
The even-parity vector potential Lagrangian then simplifies to
\begin{eqnarray}
L_{{\rm EM},(e)}
&=& \sum_{\ell,m} \ell(\ell+1) \int_{-\infty}^\infty  \Bigl\{ 
\frac{\ell(\ell+1)(1-2M/r)}{2r^2} |{\cal H}_\ell^{-1}\dot Z_{\ell,m}|^2
+   \frac12 |({\cal H}_\ell^{-1}\dot Z_{\ell,m})'|^2
-\frac12 |Z_{\ell,m}|^2
\Bigr\}\,dr_\star
\nonumber \\
&=& \sum_{\ell,m} \ell(\ell+1) \int_{-\infty}^\infty  \Bigl\{ 
 \frac12 ({\cal H}_\ell^{-1}\dot Z_{\ell,m})^\ast\left[ -\partial^2_{r_\star} + \frac{\ell(\ell+1)(1-2M/r)}{r^2} \right]({\cal H}_\ell^{-1}\dot Z_{\ell,m})
-\frac12 |Z_{\ell,m}|^2
\Bigr\}\,dr_\star
\nonumber \\
&=& \sum_{\ell,m} \ell(\ell+1) \int_{-\infty}^\infty  \Bigl\{ 
 \frac12 \dot Z_{\ell,m}^\ast{\cal H}_\ell^{-1}\dot Z_{\ell,m}
-\frac12 |Z_{\ell,m}|^2
\Bigr\}\,dr_\star,
\label{eq:Ao-2}
\end{eqnarray}
where the second line requires us to use integration by parts on $|({\cal H}_\ell^{-1}\dot Z_{\ell,m})'|^2$ and combine the first two terms, and then in the last line we recognized that the operator in brackets is ${\cal H}_\ell$, and then used ${\cal H}_\ell^{-1\,\dagger} {\cal H}_\ell{\cal H}_\ell^{-1} = {\cal H}_\ell^{-1}$.

For both the even and odd-parity sectors, the key is to find the eigenfunctions of the ${\cal H}_\ell$ operator. These are the usual mode functions of the electromagnetic field. The eigenfunctions are denoted $\Psi_{\ell,\omega}(r_\star)$, satisfying ${\cal H}_\ell\Psi_{\ell,\omega} = \omega^2\Psi_{\ell,\omega}$. Since ${\cal H}_\ell$ is a second-order differential operator, there are actually two solutions; we can see that at $r_\star\rightarrow\pm\infty$ these must be linear combinations of $e^{i\omega r_\star}$ and $e^{-i\omega r_\star}$. We choose the basis of scattering solutions
\begin{equation}
\Psi_{{\rm in},\ell,\omega}(r_\star) \rightarrow
\left\{ \begin{array}{lcl} T_{1,\ell,\omega} e^{-i\omega r_\star} & ~ & r_\star\rightarrow -\infty \\
e^{-i\omega r_\star} + R_{1,\ell,\omega} e^{i\omega r_\star} & & r_\star\rightarrow\infty
\end{array}\right.
\label{eq:psi-in}
\end{equation}
and
\begin{equation}
\Psi_{{\rm up},\ell,\omega}(r_\star) \rightarrow
\left\{ \begin{array}{lcl} e^{i\omega r_\star} - R_{1,\ell,\omega}^\ast e^{2i\arg T_{1,\ell,\omega}} e^{-i\omega r_\star} & ~ & r_\star\rightarrow -\infty \\
T_{1,\ell,\omega} e^{i\omega r_\star} & & r_\star\rightarrow\infty
\end{array}\right.,
\label{eq:psi-up}
\end{equation}
where $|R_{1,\ell,\omega}|^2 + |T_{1,\ell,\omega}|^2 =1$ (by matching of the Wronskian relation for $\{ \Psi_{{\rm in},\ell,\omega}, \Psi_{{\rm in},\ell,\omega}^\ast \}$) and the coefficients in Eq.~(\ref{eq:psi-in}) come from the matching of the Wronskian relations for $\{ \Psi_{{\rm up},\ell,\omega}, \Psi_{{\rm in},\ell,\omega} \}$ and $\{ \Psi_{{\rm up},\ell,\omega}, \Psi_{{\rm in},\ell,\omega}^\ast \}$. One recognizes $R_{1,\ell,\omega}$ and $T_{1,\ell,\omega}$ as the usual reflection and transmission coefficients from 1D scattering theory.
These satisfy the inner product relations
\begin{equation}
(\Psi_{X,\ell,\omega} | \Psi_{X',\ell,\omega'})_{r_\star} =
 2\pi \delta(\omega-\omega') \delta_{ XX'},
\end{equation}
where $X$ and $X'$ represent either ``in'' or ``up.'' To compute emission as seen by a distant observer, we will also need the ``out'' and ``down'' basis, which is related to the \{in,up\} basis as described in Appendix~\ref{app:out-down}.

\subsection{The quantized electromagnetic potentials in the generalized Coulomb gauge}

The usual quantization procedure can then be used for Eq.~(\ref{eq:Ao-1}):
\begin{equation}
A_{(o),\ell,m}(r_\star) = \frac1{\sqrt{\ell(\ell+1)}} \int_0^\infty \frac{d\omega}{2\pi} \,\frac1{\sqrt{2\omega}}
\sum_{ X\in\{\rm in,up\}}
\Bigl[ \Psi_{X,\ell,\omega}(r_\star) \hat a_{X,\ell,m,\omega,(o)}
+ (-1)^m\Psi_{X,\ell,\omega}^\ast(r_\star) \hat a_{X,\ell,-m,\omega,(o)}^\dagger \Bigr],
\label{eq:Ao-exp}
\end{equation}
with the commutation relations
\begin{equation}
[\hat a_{X,\ell,m,\omega,(o)},\hat a_{X',\ell,m,\omega',(o)}^\dagger] = 2\pi\delta(\omega-\omega')\delta_{ XX'};
\end{equation}
and the
Hamiltonian
\begin{equation}
H_{{\rm EM},(o)} = \sum_{\ell,m} \int_0^\infty \frac{d\omega}{2\pi}\,\omega \sum_{ X\in\rm\{in,up\}}
 \hat a_{X,\ell,m,\omega,(o)}^\dagger\hat a_{X,\ell,m,\omega,(o)}.
\label{eq:HEMo}
\end{equation}
The $(-1)^m$ and the $-m$ in Eq.~(\ref{eq:Ao-exp}) result from the relation for a real variable expressed in complex spherical harmonics: if $f$ is real, then $f_{\ell,m} = (-1)^mf^\ast_{\ell,-m}$. A similar procedure applies to Eq.~(\ref{eq:Ao-2}) for the even parity sector:
\begin{equation}
Z_{\ell,m}(r_\star) = \frac1{\sqrt{\ell(\ell+1)}} \int_0^\infty \frac{d\omega}{2\pi} \,\sqrt{\frac\omega 2} \sum_{X\in\rm\{in,up\}}
\Bigl[ \Psi_{X,\ell,\omega}(r_\star) \hat a_{X,\ell,m,\omega,(e)}
+ (-1)^m\Psi_{X,\ell,\omega}^\ast(r_\star) \hat a_{X,\ell,-m,\omega,(e)}^\dagger \Bigr];
\label{eq:Ae-exp}
\end{equation}
note that the normalization has an extra $\omega$ is in the numerator because this time there is an ${\cal H}_\ell^{-1}$ in the kinetic term instead of an ${\cal H}_\ell$ in the potential term. Finally we arrive at the expressions for the vector potential:
\begin{eqnarray}
A_{r_\star} &=& \frac{1-2M/r}{r^2} \sum_{\ell=1}^\infty \sum_{m=-\ell}^\ell 
\sqrt{\ell(\ell+1)} \int_0^\infty \frac{d\omega}{2\pi} \,\frac1{\sqrt{2\omega^3}}
\nonumber \\ && \times
\sum_{X\in\rm\{in,up\}}\Bigl[ 
\Psi_{X,\ell,\omega}(r_\star) \hat a_{X,\ell,m,\omega,(e)}
+ (-1)^m\Psi_{X,\ell,\omega}^\ast(r_\star) \hat a_{X,\ell,-m,\omega,(e)}^\dagger \Bigr]
Y_{\ell,m}(\theta,\phi)
~~~~{\rm and}
\nonumber \\
A_{\theta} \pm \frac{i}{\sin\theta} A_\phi &=& \mp \sum_{\ell=1}^\infty \sum_{m=-\ell}^\ell
\int_0^\infty \frac{d\omega}{2\pi} \,\frac1{\sqrt{2\omega}}\sum_{X\in\rm\{in,up\}} \Bigl\{
\frac1\omega
 \Psi'_{X,\ell,\omega}(r_\star) \hat a_{X,\ell,m,\omega,(e)}
+ (-1)^m\frac1\omega \Psi'{^\ast_{\!X,\ell,\omega}}(r_\star) \hat a_{X,\ell,-m,\omega,(e)}^\dagger
\nonumber \\ &&
 \pm i
\Bigl[ \Psi_{X,\ell,\omega}(r_\star) \hat a_{X,\ell,m,\omega,(o)}
+ (-1)^m\Psi_{X,\ell,\omega}^\ast(r_\star) \hat a_{X,\ell,-m,\omega,(o)}^\dagger \Bigr]
 \Bigr\}\, {_{\pm 1}}Y_{\ell,m}(\theta,\phi).
 \label{eq: vector_potential_final}
\end{eqnarray}

\section{The electron field}
\label{sec: electron_field}

We now investigate the electron field. We follow the classical solution to the Dirac equation by \citet{1957RvMP...29..465B}. The Dirac field was quantized on the Schwarzschild spacetime by \citet{1975PhRvD..12..350B}, although there the mode operators are not explicitly written in terms of annihilation and creation operators. \citet{2013PhRvD..87f4027C} perform a quantization similar to what is done here, but for the massless case.

\subsection{Angular operator}

The angular operator in the separation of the Dirac equation (Eq.~35 of \cite{1957RvMP...29..465B}) is
\begin{equation}
K = \left( \begin{array}{cccc}
-i & 0 & 0 & 0 \\
0 & i & 0 & 0 \\
0 & 0 & i & 0 \\
0 & 0 & 0 & -i
\end{array} \right) \partial_\theta
+
\left( \begin{array}{cccc}
0 & 1 & 0 & 0 \\
-1 & 0 & 0 & 0 \\
0 & 0 & 0 & -1 \\
0 & 0 & 1 & 0
\end{array} \right) \frac{\partial_\phi}{\sin\theta}.
\end{equation}
It has eigenvalues $k = \pm1,\pm2,\pm3,...$, each of which is repeated. We express the eigenfunctions in modern notation using the spin-weighted spherical harmonics \cite{1967JMP.....8.2155G}:
\begin{equation}
\Theta^{({ F})}_{k,m} =
\frac{\sqrt{\sin\theta}}{2}
\left( \begin{array}{c}
{_{\frac12}}Y_{j,m}(\theta,\phi) - is \,{_{-\frac12}}Y_{j,m}(\theta,\phi) \\
-{_{\frac12}}Y_{j,m}(\theta,\phi) - is \,{_{-\frac12}}Y_{j,m}(\theta,\phi) \\
0 \\ 0
\end{array}\right)
~~{\rm and}~~
\Theta^{({ G})}_{k,m} =
\frac{\sqrt{\sin\theta}}{2}
\left( \begin{array}{c}
0 \\ 0 \\
-{_{\frac12}}Y_{j,m}(\theta,\phi) - is \,{_{-\frac12}}Y_{j,m}(\theta,\phi) \\
{_{\frac12}}Y_{j,m}(\theta,\phi) - is \,{_{-\frac12}}Y_{j,m}(\theta,\phi)
\end{array}\right),
\label{eq:TFGMODE}
\end{equation}
where $j = |k|-\frac12$, $m$ ranges from $-j ... +j$, $s=k/|k|$ is the sign of $k$, and $(F)$ or $(G)$ indicates which of the two eigenfunctions is under consideration.
There are two linearly independent eigenfunctions for each $j$ and $m$, which is expected since the $K$ operator breaks into two equivalent $2\times 2$ blocks.
This form of the angular eigenfunctions can be verified directly using the spin raising and lowering operators:
\begin{equation}
 \partial_\theta \left[ \sqrt{\sin\theta}\,{_{\pm\frac12}}Y_{j,m}(\theta,\phi) \right]
\mp i \frac{\partial_\phi}{\sin\theta} \left[ \sqrt{\sin\theta}\,{_{\pm\frac12}}Y_{j,m}(\theta,\phi) \right]
= \pm\left(j+\frac12\right)\sqrt{\sin\theta} \,{_{\mp\frac12}}Y_{j,m}(\theta,\phi).
\end{equation}
These eigenfunctions satisfy the orthonormality relation
\begin{eqnarray}
\int_0^{2\pi}\int_0^\pi \Theta_{km}^{(F)\dagger}(\theta,\phi) \Theta^{(F)}_{k'm'}(\theta,\phi)\,d\theta\,d\phi &=&
\delta_{kk'} \delta_{mm'},
\nonumber \\
\int_0^{2\pi}\int_0^\pi \Theta_{km}^{(G)\dagger}(\theta,\phi) \Theta^{(G)}_{k'm'}(\theta,\phi)\,d\theta\,d\phi &=& 
\delta_{kk'} \delta_{mm'}, ~~~~{\rm and}
\nonumber \\
\int_0^{2\pi}\int_0^\pi \Theta_{km}^{(F)\dagger}(\theta,\phi) \Theta^{(G)}_{k'm'}(\theta,\phi)\,d\theta\,d\phi &=&
0.
\label{eq:norm-ang}
\end{eqnarray}

\subsection{The radial modes}

The solutions of the Dirac equation $(\tilde{\boldsymbol\gamma}^\mu\nabla_\mu - m)\psi = 0$ are of the form
\begin{equation}
\psi(r,\theta,\phi,t) = \frac{1}{r(1-2M/r)^{1/4}\sqrt{\sin\theta}} [F(r) \Theta^{(F)}_{k,m}(\theta,\phi) + G(r) \Theta^{(G)}_{k,m}(\theta,\phi) ] e^{-i h  t},
\end{equation}
where $ h $ is the energy at $r\rightarrow\infty$ and the radial functions $F$ and $G$ satisfy Eq.~(39) of \citet{1957RvMP...29..465B}\footnote{We have corrected a sign error in $\mu$ in the original reference.}:
\begin{equation}
 h  \left( \begin{array}cF \\ G\end{array}\right)
= 
\left( \begin{array}{cc} \mu\sqrt{1-\frac{2M}r} & \frac kr\sqrt{1-\frac{2M}r} + \partial_{r_\star} \\ \frac kr\sqrt{1-\frac{2M}r} -\partial_{r_\star} & -\mu\sqrt{1-\frac{2M}r} \end{array}\right)
\left( \begin{array}cF \\ G\end{array}\right)
\label{eq:FG-DE}
\end{equation}
(we have re-written this in terms of derivatives with respect to $r_\star$). Note that the $2\times 2$ matrix is a Hermitian operator with respect to the usual inner product. Since this is a coupled system of 2 first-order ordinary differential equations, there are 2 linearly independent solutions. Note that these are real equations, so if $(F,G)$ is a complex solution then so is $(F^\ast,G^\ast)$. Also we note that if one flips the sign of $ h $ and $k$ and swaps $F\leftrightarrow G$, then one again has a solution, so we need only construct the solutions for $ h >0$.

We are interested in the limiting solutions of this near the horizon and at spatial infinity. Near the horizon ($r_\star \rightarrow -\infty$), the terms without $\sqrt{1-2M/r}$ are dominant and we have the two linearly independent solutions
\begin{equation}
\left( \begin{array}c F \\
G \end{array} \right)
\propto \left( \begin{array}c \sqrt{ h } \\ -i\sqrt{ h } \end{array} \right) e^{i h  r_\star}
~~~{\rm and}~~~
\left( \begin{array}c F \\
G \end{array} \right)
\propto \left( \begin{array}c \sqrt{ h } \\ i\sqrt{ h } \end{array} \right) e^{-i h  r_\star}.
\end{equation}
At spatial infinity ($r_\star\rightarrow\infty$), the $k/r$ terms drop out, and $1-2M/r\rightarrow 1$. The wavenumber in the Wentzel-Kramers-Brillouin approximation is $p \approx \sqrt{ h ^2-\mu^2} + \zeta/r_\star + ...$, where the coefficient of the leading-order correction is
\begin{equation} 
\zeta = \frac{\mu^2M}{\sqrt{ h ^2-\mu^2}}.
\end{equation}
The phase of the wave function is $\int p\,dr_\star = \sqrt{ h ^2-\mu^2}\,r_\star + \zeta\ln (r_\star/2M) + ...$; note that there is a logarithmically divergent phase due to the gravitational potential, which is analogous to the similar phenomenon in the Coulomb wave functions and generically occurs with long-range $1/r$ potentials (see, e.g., \S14.5 of Ref.~\cite{1965hmfw.book.....A}). Including this phase in the large-$r$ limit, we get:
\begin{equation}
\left( \begin{array}c F \\
G \end{array} \right)
\propto \left( \begin{array}c \sqrt{ h +\mu} \\ -i\sqrt{ h -\mu} \end{array} \right) e^{i\zeta \ln (r_\star/2M)} e^{i\sqrt{ h ^2-\mu^2}\, r_\star}
~~~{\rm and}~~~
\left( \begin{array}c F \\
G \end{array} \right)
\propto \left( \begin{array}c \sqrt{ h +\mu} \\ i\sqrt{ h -\mu} \end{array} \right) e^{-i\zeta\ln (r_\star/2M)} e^{-i\sqrt{ h ^2-\mu^2}\, r_\star}.
\label{eq: large_r_scatter_soln}
\end{equation}
Note that there is an oscillatory solution here only if $| h |>\mu$; otherwise the solutions become evanescent.

We now construct scattering solutions, constrained by the Wronskian relation for Eq.~(\ref{eq:FG-DE}), i.e., that if there are two solutions $(F_1,G_1)$ and $(F_2,G_2)$ then $F_1G_2-F_2G_1$ is independent of $r_\star$. For $ h >\mu$, there is an ``ingoing'' solution
\begin{equation}
\left( \begin{array}c F_{\rm in} \\
G_{\rm in} \end{array} \right)
\rightarrow
\left\{ \begin{array}{lll}
v^{-1/2} \left( \begin{array}c \sqrt{ h +\mu} \\ i\sqrt{ h -\mu} \end{array} \right) e^{-i\zeta \ln (r_\star/2M)}  e^{-i\sqrt{ h ^2-\mu^2}\, r_\star} \\
~~~~+ v^{-1/2} R_{\frac12,k, h } \left( \begin{array}c \sqrt{ h +\mu} \\ -i\sqrt{ h -\mu} \end{array} \right) e^{i\zeta \ln (r_\star/2M)}  e^{i\sqrt{ h ^2-\mu^2}\, r_\star}  & & r_\star\rightarrow+\infty \\
 T_{\frac12,k, h } \left( \begin{array}c \sqrt{ h } \\ i\sqrt{ h } \end{array} \right) e^{-i h  r_\star} & & r_\star\rightarrow-\infty
\end{array}\right.~~( h >\mu),
\label{eq:in-e.gt}
\end{equation}
with $|R_{\frac12,k, h }|^2 + |T_{\frac12,k, h }|^2 = 1$ since $(F_{\rm in},G_{\rm in})$ and $(F_{\rm in}^\ast,G_{\rm in}^\ast)$ satisfy the Wronskian relations. We have defined
\begin{equation}
v = \frac{\sqrt{ h ^2-\mu^2}} h ,
\end{equation}
which is also equal to the velocity of the electron at $\infty$.
There is also an ``upgoing'' solution,
\begin{equation}
\left( \begin{array}c F_{\rm up} \\
G_{\rm up} \end{array} \right)
\rightarrow
\left\{ \begin{array}{lll}
\left( \begin{array}c \sqrt{ h } \\ -i\sqrt{ h } \end{array} \right) e^{i h  r_\star}
- R_{\frac12,k, h }^\ast e^{2i \arg T_{1/2,k, h }} \left( \begin{array}c \sqrt{ h } \\ i\sqrt{ h } \end{array} \right) e^{-i h  r_\star} & & r_\star\rightarrow-\infty \\
T_{\frac12,k, h } v^{-1/2} \left( \begin{array}c \sqrt{ h +\mu} \\ -i\sqrt{ h -\mu} \end{array} \right) e^{i\zeta \ln (r_\star/2M)}  e^{i\sqrt{ h ^2-\mu^2}\, r_\star}  & & r_\star\rightarrow+\infty
\end{array}\right.~~( h >\mu),
\label{eq:up-gt}
\end{equation}
where the Wronskian rules have been used to find the coefficients. For the case of $ h <\mu$, there are only ``up'' modes (there is no ``in''):
\begin{equation}
\left( \begin{array}c F_{\rm up} \\
G_{\rm up} \end{array} \right)
\rightarrow
\left\{ \begin{array}{lll}
\left( \begin{array}c \sqrt{ h } \\ -i\sqrt{ h } \end{array} \right) e^{i h  r_\star}
+ e^{2i\delta_{1/2,k, h }} \left( \begin{array}c \sqrt{ h } \\ i\sqrt{ h } \end{array} \right) e^{-i h  r_\star} & & r_\star\rightarrow-\infty \\
{\rm exponentially~decaying} & & r_\star\rightarrow+\infty
\end{array}\right.~~(0< h <\mu),
\label{eq:up-lt}
\end{equation}
where the reflection coefficient must have unit modulus, and the phase is contained in $\delta_{1/2,k, h }$. 
We can further define the negative energy modes by $F_{X,-k,- h } = G^\ast_{X,k, h }$ and $G_{X,-k,- h } = F^\ast_{X,k, h }$, which satisfy the same scattering wave boundary conditions.
These modes satisfy
\begin{equation}
\int_{-\infty}^\infty (F_{X,k, h }^\star F_{X',k, h '} + G_{X,k, h }^\star G_{X',k, h '})\,dr_\star = 
4\pi\left| h \right|\,\delta_{XX'}\delta( h - h '),
\label{eq:FFGG}
\end{equation}
where $X,X'\in\{{\rm in,up}\}$ (one proves this using $v^{-1}\,d h  = d\sqrt{ h ^2-\mu^2}$).

\subsection{Quantization}

The next step is to find the canonical normalization of the modes. The Dirac Lagrangian is
\begin{equation}
L_{\rm Dirac} = \frac{dS_{\rm Dirac}}{dt} = -\int \psi^\dagger{\boldsymbol\beta} \left[
\frac{-i {\boldsymbol\beta} \nabla_t 
+ \tilde{\boldsymbol\gamma}_1 \nabla_{r_\star}}{\sqrt{1-2M/r}}
+ \frac1r \left( \tilde{\boldsymbol\gamma}_2 \nabla_{\theta}
+ \frac1{\sin\theta} \tilde{\boldsymbol\gamma}_3 \nabla_{\phi} \right)
+ \mu \right] \psi\, r^2\left( 1 - \frac{2M}r \right)\,dr_\star\,\sin\theta\,d\theta\,d\phi,
\end{equation}
and it follows from the time-derivative term that the canonical anticommutation relation is
\begin{equation}
\{ \psi_A(r_\star,\theta,\phi), \psi_B^\dagger(r'_\star,\theta',\phi') \}
= \frac{\delta_{AB} \delta(r_\star-r'_\star) \delta(\theta-\theta') \delta(\phi-\phi')}{r^2(1-2M/r)^{1/2}\sin\theta} .
\end{equation}
Using the radial mode normalization, Eq.~(\ref{eq:FFGG}), and the angular mode normalization of Eq.~(\ref{eq:norm-ang}), we conclude that the proper normalization is
\begin{eqnarray}
\label{eq:fermion_field}
\psi(r,\theta,\phi) &=& \int_0^\infty \frac{d h }{2\pi} \frac1{\sqrt{2 h }} \sum_{Xkm}
\frac{1}{r(1-2M/r)^{1/4}\sqrt{\sin\theta}} \Bigl\{ [F_{X,k, h }(r) \Theta^{(F)}_{k,m}(\theta,\phi) + G_{X,k, h }(r) \Theta^{(G)}_{k,m}(\theta,\phi) ] \hat b_{X,k,m, h } 
\nonumber \\ &&
+
[G^\ast_{X,-k, h }(r) \Theta^{(F)}_{k,m}(\theta,\phi) + F^\ast_{X,-k, h }(r) \Theta^{(G)}_{k,m}(\theta,\phi) ] \hat d^\dagger_{X,k,m, h }
\Bigr\}
\label{eq: psi_electron}
\end{eqnarray}
with the anticommutation relation
\begin{equation}
\{ \hat b_{X,k,m, h }, \hat b^\dagger_{X',k',m', h '} \} = \{ \hat d_{X,k,m, h }, \hat d^\dagger_{X',k',m', h '} \} 
= 2\pi\,\delta( h - h ') \delta_{XX'} \delta_{kk'} \delta_{mm'}.
\end{equation}
The Hamiltonian, as per the usual procedure, is
\begin{equation}
H_{\rm Dirac} = \int_0^\infty \frac{d h }{2\pi}\, h  \sum_{Xkm} \left( \hat b^\dagger_{X,k,m, h } \hat b_{X,k,m, h } 
+ \hat d^\dagger_{X,k,m, h } \hat d_{X,k,m, h }  \right).
\label{eq:H-Dirac}
\end{equation}

\section{Interaction Hamiltonian}
\label{sec: H_int}

We are now interested in the interaction Hamiltonian, $H_{\rm int}$. There are two steps to this: the canonical construction of the interaction Hamiltonian and evaluation of the angular integrals (\S\ref{ss:H_int_canonical}); and the expression in terms of overlap integrals and operators that is useful for time-dependent perturbation theory (\S\ref{ss:I}).

\subsection{Canonical construction of $H_{\rm int}$}
\label{ss:H_int_canonical}

Since no time derivatives of the fields appear explicitly in the action, we have $H_{\rm int} = -L_{\rm int} = -dS_{\rm int}/dt$. Just like in flat spacetime QED, the interaction comes from the $-ieA_\mu$ term in the covariant derivative (Eq.~\ref{eq:covariant-derivative}) substituted into the action (Eq.~\ref{eq:action-QED}).
We can thus write interaction Hamiltonian between the electron and photon vertices as
\begin{equation}
\label{eq: Hamiltonian_interact}
H_{\rm int}
= -ie \int \sqrt{-g}\,d^3x\, \overline{\psi}(x)\gamma^{\mu}A_{\mu}(x)\psi(x),
\end{equation}
where the $-i$ is needed since this convention for metric implies that the space components of $\boldsymbol{\gamma}^{\rm i}$ and  $\boldsymbol{\beta}$  are hermitian, the volume element $\sqrt{-g}$ of the Schwarzschild metric in the tortoise coordinate system is given by $(1-2M/r)r^2\sin{\theta}$, and the functions for $A_{\mu}(x)$ and $\psi(x)$ are given by Eqs.~(\ref{eq: vector_potential_final}) and (\ref{eq: psi_electron}), respectively.

We can expand the interaction Hamiltonian to give 
\begin{eqnarray}
H_{\rm int} &=& -ie \int \sqrt{1 - \frac{2M}r} \,dr_{*}\,d\theta \,d\phi \int \frac{dh \,dh'}{(2\pi)^2 2\sqrt{hh'}} \sum_{X_{s}} \sum_{X_{s'}}~ \Bigl[\,\overline{(F_{X,k, h }(r) \Theta^{(F)}_{k,m}(\theta,\phi) + G_{X,k, h }(r) \Theta^{(G)}_{k,m}(\theta,\phi))}\, \hat b^\dagger_{X,k,m, h } 
\nonumber \\  &&
~~~~+\,
\overline{(G^\ast_{X,-k, h }(r) \Theta^{(F)}_{k,m}(\theta,\phi) + F^\ast_{X,-k, h }(r) \Theta^{(G)}_{k,m}(\theta,\phi))}\, \hat d_{X,k,m, h }\Bigr]
\nonumber \\ && \times
\Bigl[\Bigl(1-\frac{2M}r\Bigr)^{-1/2}\tilde{\boldsymbol\gamma}^1A_{r_{*}} + \frac{1}{2r}(\tilde{\boldsymbol\gamma}^2 + i\tilde{\boldsymbol\gamma}^3)\Bigl(A_{\theta} -\frac{i}{\sin{\theta}}A_{\phi}\Bigr)
+ \frac{1}{2r}(\tilde{\boldsymbol\gamma}^2 - i\tilde{\boldsymbol\gamma}^3)\Bigl(A_{\theta} + \frac{i}{\sin{\theta}}A_{\phi}\Bigr)\Bigr]
\nonumber \\ && \times
\Bigl[(F_{X',k', h' }(r) \Theta^{(F)}_{k',m'}(\theta,\phi) + G_{X',k', h' }(r) \Theta^{(G)}_{k',m'}(\theta,\phi)) \hat b_{X',k',m', h' } 
\nonumber \\ &&
~~~~+
(G^\ast_{X',-k', h' }(r) \Theta^{(F)}_{k',m'}(\theta,\phi) + F^\ast_{X',-k', h' }(r) \Theta^{(G)}_{k',m'}(\theta,\phi)) \hat d^\dagger_{X',k',m', h' }\Bigr],
\label{eq: Hamiltonian_interact_expand}
\end{eqnarray}
where the overbar terms are the usual quantities hermitian conjugated and multiplied from the right by $\boldsymbol{\beta}$. The matrices ${\boldsymbol\gamma}^i$ correspond to the flat spacetime matrices, $\tilde{\boldsymbol\gamma}^i$ in Eq. (\ref{eq: gamma_matrices}) with the appropriate tetrad transformation. The first term in this expansion ($A_{r_*}$ term) can be simplified as
\begin{eqnarray}
H_{\rm int}^{r_*} &=& -ie\int dr_* \frac{dhdh'd\omega}{(2\pi)^3 2\sqrt{hh'}}\frac{1}{\sqrt{2\omega^3}}\frac{1-2M/r}{r^2}\sum_{X_{\gamma} \ell m_{\gamma}}\sqrt{\ell (\ell + 1)} \sum_{X km} \sum_{X'k'm'}
\nonumber \\ &&
\Bigl[(F^*_{Xkh}F^*_{X'-k'h'} - G^*_{Xkh}G^*_{X'-k'h'})\Psi_{X_\gamma \ell \omega} \Delta^{kk'l}_{mm'm_\gamma} \hat b^\dagger_{Xkmh} \hat a_{X_{\gamma} \ell m \omega, (e)} \hat d^\dagger_{X'k'm'h'} 
\nonumber \\ &&
+ (G_{X-kh}F^*_{X'-k'h'} - F_{X-kh}G^*_{X'-k'h'})\Psi_{X_\gamma \ell \omega} \Delta^{kk'l}_{mm'm_\gamma} \hat d_{Xkmh} \hat a_{X_{\gamma} \ell m \omega, (e)} \hat d^\dagger_{X'k'm'h'}
\nonumber\\ &&
+ (F^*_{Xkh}G_{X'k'h'} - G^*_{Xkh}F_{X'k'h'})\Psi_{X_\gamma \ell \omega} \Delta^{kk'l}_{mm'm_\gamma} \hat b^\dagger_{Xkmh} \hat a_{X_{\gamma} \ell m \omega, (e)} \hat b_{X'k'm'h'}
\nonumber\\ &&
+ (G_{X-kh}G_{X'k'h'} - F_{X-kh}F_{X'k'h'})\Psi_{X_\gamma \ell \omega} \Delta^{kk'l}_{mm'm_\gamma} \hat d_{Xkmh} \hat a_{X_{\gamma} \ell m \omega, (e)} \hat b_{X'k'm'h'} \Bigr]
+ {\rm h.c.},
\label{eq: H_int_firstterm}
\end{eqnarray}
where the angular integral is
\begin{equation}
\Delta^{kk'\ell}_{mm'm_\gamma} \equiv \frac{(-1)^{m + 1/2} }{2}\sqrt{\frac{(2j + 1)(2j' + 1)(2\ell + 1)}{4\pi}} \tj{j}{j'}{\ell}{-m}{m'}{m_{\gamma}} \tj{j}{j'}{\ell}{\frac12}{-\frac12}{0} (1 + ss'(-1)^{j - j' + \ell}).
\label{eq: DELTA}
\end{equation}
The angular expression in Eq.~(\ref{eq: DELTA}) has a set of selection rules between the transition of spinor from $j \rightarrow j'$ with emission/absorption of a photon of angular momentum $\ell$. We can see Eq.~(\ref{eq: DELTA}) is non-zero for two cases: (1) when the sign of the angular momentum of the two spinor states are the same ($s=s'$) then only \textit{even} transitions are allowed i.e. $j-j'$ and $\ell$ must be either even or odd, and (2) when the sign of the angular momentum are different $s \neq s'$, then $j-j'$ and $\ell$ must have different types i.e. \textit{odd} transitions.

We then compute the $H^{\theta \pm i\phi}_{\rm int}$ terms corresponding to the second and third term in the expansion of ${\boldsymbol\gamma}^\mu A_{\mu}$:
\begin{eqnarray}
H^{\theta, \phi}_{\rm int} &=& -ie \int \frac{\sqrt{1 - 2M/r}}{r} dr_* \frac{dh dh'd\omega}{4(2\pi)^3 \sqrt{2\omega} \sqrt{hh'}} \sum_{X_{\gamma} \ell m_{\gamma}}
\sum_{X km} \sum_{X'k'm'}
\nonumber \\ &&
\Bigl[(F^*_{Xkh}G_{X'k'h'} + G^*_{Xkh}F_{X'k'h'}) \Pi^{kk'l\,-}_{mm'm_\gamma} \frac{1}{\omega} \Psi'_{X_{\gamma} \ell \omega} \hat b^\dagger_{Xkmh} \hat a_{X_{\gamma}, \ell, \omega, m, (e)} \hat b_{X'k'm'h'} - i \omega \rm{(e \leftrightarrow o, \Pi^{- \leftrightarrow +},\dstrike)} 
\nonumber \\ &&
+ (G_{X-kh}G_{X'k'h'} + F_{X-kh}F_{X'k'h'})\Pi^{kk'l\,-}_{mm'm_\gamma} \frac{1}{\omega} \Psi'_{X_{\gamma} \ell \omega} \hat d_{Xkmh} \hat a_{X_{\gamma}, \ell, \omega, m,(e)} \hat b_{X'k'm'h'} - i \omega \rm{(e \leftrightarrow o, \Pi^{- \leftrightarrow +},\dstrike)}
\nonumber \\ &&
+ (F^*_{Xkh}F^*_{X'-k'h'} + G^*_{Xkh}G^*_{X'-k'h'})\Pi^{kk'l\,-}_{mm'm_\gamma} \frac{1}{\omega} \Psi'_{X_{\gamma} \ell \omega} \hat b^\dagger_{Xkmh} \hat a_{X_{\gamma}, \ell, \omega, m,(e)} \hat d^\dagger_{X'k'm'h'} - i \omega \rm{(e \leftrightarrow o, \Pi^{- \leftrightarrow +},\dstrike)}
\nonumber \\ &&
+ (G_{X-kh}F^*_{X'-k'h'} + F_{X-kh}G^*_{X'-k'h'})\Pi^{kk'l\,-}_{mm'm_\gamma} \frac{1}{\omega} \Psi'_{X_{\gamma} \ell \omega} \hat d_{Xkmh} \hat a_{X_{\gamma}, \ell, \omega, m,(e)} \hat d^\dagger_{X'k'm'h'} - i \omega \rm{(e \leftrightarrow o, \Pi^{- \leftrightarrow +},\dstrike)}\Bigr]
\nonumber \\ && + {\rm h.c.},
\label{eq: H_int_secondterm}
\end{eqnarray}
where the angular integral is:
\begin{equation}
\Pi^{kk'\ell\,\pm}_{mm'm_\gamma} \equiv s (-1)^{m - 1/2} \sqrt{\frac{(2j +1)(2j' + 1)(2\ell + 1)}{4\pi}} \tj{j}{j'}{\ell}{-m}{m'}{m_{\gamma}} \tj{j}{j'}{\ell}{-\frac12}{-\frac12}{1} (1 \mp ss'(-1)^{j - j' + \ell}),
\label{eq: PI}
\end{equation}
Here the terms labeled $-i \omega (\rm{e \leftrightarrow o}, \Pi^{- \leftrightarrow +}, \dstrike)$ are the odd parity sectors of the $A_{\theta} \pm iA_{\phi}/\sin{\theta}$ with the $r_*$ derivative of the $\Psi_{X_{\gamma} \ell \omega}$ removed, the angular quantity switches signs, and the prefactors of $F$ and $G$ are the same in either swapped case. 

Similarly to Eq.~(\ref{eq: DELTA}), there are a set of selection rules for Eq.~(\ref{eq: PI}). For the $\Pi^{kk'\ell -}_{mm'm_{\gamma}}$, these have the same selection rules as Eq.~(\ref{eq: DELTA}) but have the opposite set of rules for $\Pi^{kk'\ell +}_{mm'm_{\gamma}}$. This is because the angular part of the vector potentials have both even and odd contributions from the spherical basis vectors. So the radial part of the vector potential (even symmetry) and the even sector of the angular part of the interaction Hamiltonian should have the same set of selection rules.

Some useful identities about the angular integrals are proved in Appendix~\ref{app:identities}.

\subsection{Expression in terms of 3-mode overlap integrals}
\label{ss:I}

Now that we have the interaction Hamiltonian, we will follow a similar set up as laid out in Chapter 6 of \citet{2004ASSL..307.....L}. The first step is to collect terms involving each of the photon annihilation operators (the creation operators are in the Hermitian conjugate term). Following the steps from \citet{2004ASSL..307.....L}, we first re-write the interaction terms of Eqs. (\ref{eq: H_int_firstterm}) and (\ref{eq: H_int_secondterm}) as:
\begin{eqnarray}
H_{\rm int}(t) &=& e \sum_{X_\gamma\ell m_\gamma p} \int \frac{d\omega}{2\pi}\,
\hat a_{X_{\gamma}, \ell, \omega, m_{\gamma}, (p)} \int \frac{dh\,dh'}{(2\pi)^2} \sum_{XkmX'k'm'} \Bigl[\hat b^\dagger_{Xkmh} \hat d^\dagger_{X'k'm'h'} I^{++}_{Xkm,X'k'm',X_\gamma \ell m_\gamma (p)}(h,h',\omega)
\nonumber \\ && +
\hat d_{Xkmh} \hat b_{X'k'm'h'} I^{--}_{Xkm,X'k'm',X_\gamma \ell m_\gamma (p)}(h,h',\omega)  +
\hat b^\dagger_{Xkmh} \hat b_{X'k'm'h'} I^{+-}_{Xkm,X'k'm',X_\gamma \ell m_\gamma (p)}(h,h',\omega)
\nonumber \\ &&
+ \hat d_{Xkmh} \hat d^\dagger_{X'k'm'h'} I^{-+}_{Xkm,X'k'm',X_\gamma \ell m_\gamma (p)}(h,h',\omega))
\Bigr] + {\rm h.c.},
\label{eq: Q_int}
\end{eqnarray}
where symbols such as $I^{++}_{Xkm,X'k'm',X_\gamma \ell m_\gamma (p)}(h,h',\omega)$ contain the contributions from both $H^{r_{*}}_{\rm int}$ and $H^{\theta,\phi}_{\rm int}$, and are given by
\begin{eqnarray}
I^{++}_{Xkm,X'k'm',X_\gamma \ell m_\gamma (e)}(h,h',\omega) &=& \frac{-i}{\sqrt{4hh'}}\Delta^{kk'l}_{mm'm_\gamma}\! \int_{-\infty}^\infty \Biggl[
(F^*_{Xkh}F^*_{X'-k'h'} - G^*_{Xkh}G^*_{X'-k'h'})
\Psi_{X_\gamma \ell \omega} \sqrt{\ell (\ell +1)} \frac{1-2M/r}{r^2 \sqrt{2\omega^3}}
\nonumber \\ &&
+ (F^*_{Xkh}F^*_{X'-k'h'} 
+ G^*_{Xkh}G^*_{X'-k'h'})\frac{k-k'}{\sqrt{\ell(\ell + 1)}} \frac{1}{\omega} \Psi'_{X_{\gamma} \ell \omega} \frac{\sqrt{1-2M/r}}{r \sqrt{2\omega}} \Biggr]\,dr_\star,
\nonumber \\
I^{--}_{Xkm,X'k'm',X_\gamma \ell m_\gamma (e)}(h,h',\omega) &=& \frac{-i}{\sqrt{4hh'}}\Delta^{kk'l}_{mm'm_\gamma}\!
\int_{-\infty}^\infty \Biggl[
(G_{X-kh}G_{X'k'h'} - F_{X-kh}F_{X'k'h'})
\Psi_{X_\gamma \ell \omega}  \sqrt{\ell (\ell +1)} \frac{1-2M/r}{r^2 \sqrt{2\omega^3}}
\nonumber \\ &&
+ (G_{X-kh}G_{X'k'h'} 
+ F_{X-kh}F_{X'k'h'})\frac{k-k'}{\sqrt{\ell(\ell + 1)}} \frac{1}{\omega} \Psi'_{X_{\gamma} \ell \omega} \frac{\sqrt{1-2M/r}}{r \sqrt{2\omega}} \Biggr]\,dr_\star,
\nonumber \\
I^{+-}_{Xkm,X'k'm',X_\gamma \ell m_\gamma (e)}(h,h',\omega) &=& \frac{-i}{\sqrt{4hh'}}\Delta^{kk'l}_{mm'm_\gamma}\!
\int_{-\infty}^\infty \Biggl[
(F^*_{Xkh}G_{X'k'h'} 
- G^*_{Xkh}F_{X'k'h'})
\Psi_{X_\gamma \ell \omega} \sqrt{\ell (\ell +1)} \frac{1-2M/r}{r^2 \sqrt{2\omega^3}}
\nonumber \\ &&
+ (F^*_{Xkh}G_{X'k'h'} 
+ G^*_{Xkh}F_{X'k'h'})\frac{k-k'}{\sqrt{\ell(\ell + 1)}} \frac{1}{\omega} \Psi'_{X_{\gamma} \ell \omega} \frac{\sqrt{1-2M/r}}{r \sqrt{2\omega}}\Biggr]\,dr_\star,
\nonumber \\
I^{-+}_{Xkm,X'k'm',X_\gamma \ell m_\gamma (e)}(h,h',\omega) &=& \frac{-i}{\sqrt{4hh'}}\Delta^{kk'l}_{mm'm_\gamma}\!
\int_{-\infty}^\infty \Biggl[
(G_{X-kh}F^*_{X'-k'h'} - F_{X-kh}G^*_{X'-k'h'})
\Psi_{X_\gamma \ell \omega} \sqrt{\ell (\ell +1)} \frac{1-2M/r}{r^2 \sqrt{2\omega^3}}
\nonumber \\ &&
+ (G_{X-kh}F^*_{X'-k'h'} 
+ F_{X-kh}G^*_{X'-k'h'})\frac{k-k'}{\sqrt{\ell(\ell + 1)}} \frac{1}{\omega} \Psi'_{X_{\gamma} \ell \omega} \frac{\sqrt{1-2M/r}}{r \sqrt{2\omega}} \Biggr]\,dr_\star,
\nonumber \\
I^{++}_{Xkm,X'k'm',X_\gamma \ell m_\gamma (o)}(h,h',\omega) &=& \frac{-1}{\sqrt{4hh'}}\Pi^{kk'\ell +}_{mm'm_{\gamma}}
\int_{-\infty}^\infty \Biggl[ (F^*_{Xkh}F^*_{X'-k'h'} 
+ G^*_{Xkh}G^*_{X'-k'h'}) \Psi_{X_{\gamma} \ell \omega} \frac{\sqrt{1-2M/r}}{2r \sqrt{2\omega}} \Biggr]\,dr_\star,
\nonumber \\
I^{--}_{Xkm,X'k'm',X_\gamma \ell m_\gamma (o)}(h,h',\omega) &=& \frac{-1}{\sqrt{4hh'}}\Pi^{kk'\ell +}_{mm'm_{\gamma}}
\int_{-\infty}^\infty \Biggl[
(G_{X-kh}G_{X'k'h'} 
+ F_{X-kh}F_{X'k'h'}) \Psi_{X_{\gamma} \ell \omega} \frac{\sqrt{1-2M/r}}{2r \sqrt{2\omega}}\Biggr]\,dr_\star,
\nonumber \\
I^{+-}_{Xkm,X'k'm',X_\gamma \ell m_\gamma (o)}(h,h',\omega) &=& \frac{-1}{\sqrt{4hh'}}\Pi^{kk'\ell +}_{mm'm_{\gamma}}
\int_{-\infty}^\infty \Biggl[
(F^*_{Xkh}G_{X'k'h'} 
+ G^*_{Xkh}F_{X'k'h'}) \Psi_{X_{\gamma} \ell \omega} \frac{\sqrt{1-2M/r}}{2r \sqrt{2\omega}}\Biggr]\,dr_\star, ~~{\rm and}
\nonumber \\
I^{-+}_{Xkm,X'k'm',X_\gamma \ell m_\gamma (o)}(h,h',\omega) &=& \frac{-1}{\sqrt{4hh'}}\Pi^{kk'\ell +}_{mm'm_{\gamma}}
\int_{-\infty}^\infty \Biggl[
(G_{X-kh}F^*_{X'-k'h'} 
+ F_{X-kh}G^*_{X'-k'h'}) \Psi_{X_{\gamma} \ell \omega} \frac{\sqrt{1-2M/r}}{2r \sqrt{2\omega}}\Biggr]\,dr_\star.
\nonumber \\ &&
\label{eq: coupling_int}
\end{eqnarray}
where we used the identities in Appendix~\ref{app:identities} to relate the angular quantities in the even sector. The interactions associated with the coupling integrals in Eq.~(\ref{eq: coupling_int}) are on a curved spacetime, but we show in App.~(\ref{app: test_case}) these expressions converge to the standard electric and magnetic dipole transitions. 

All of these integrals depend on the $m,m',m_\gamma$ quantum numbers only through the $3j$ symbols in $\Delta^{kk'\ell}_{mm'm_\gamma}$ and $\Pi^{kk'\ell\,\pm}_{mm'\gamma}$ and the $(-1)^m$; thus we define ``double-barred'' versions of the $I$'s,
\begin{equation}
I^{\tau\tau'}_{Xkm,X'k'm',X_\gamma\ell m_\gamma(p)}(h,h',\omega)
= \llbracket I^{\tau\tau'}_{Xk,X'k',X_\gamma\ell(p)}(h,h',\omega)
 \rrbracket
(-1)^{\ell-j'-m} \tj{j}{j'}{\ell}{-m}{m'}{m_\gamma}
\label{eq:double-bracket}
\end{equation}
(where $\tau,\tau'\in\{+,-\}$ and the phase $(-1)^{\ell -j' -m}$ is a standard convention)
in analogy to the double-barred matrix elements of the Wigner-Eckart theorem. 

\section{Evolution of the photon density matrix}
\label{sec: photon_evolve}

Our ultimate aim is to determine the statistical properties of the emitted radiation. In this section, we will define the initial conditions for the phase space densities of the spinors and radiation of a black hole in vacuum, i.e. no external spinor or radiation fields. Then we follow by expressing the interaction Hamiltonian in Eq.~(\ref{eq: Hamiltonian_interact}) in the interaction picture, in order to compute the evolution of the phase space density for the emitted radiation (following a similar procedure as laid out in Chap.~6 of \citet{2004ASSL..307.....L}).

\subsection{Occupation functions and outgoing radiation}

We define the occupation function for the photons (phase space densities) as
\begin{equation}
\label{eq: phase_density_photonI}
\langle \hat a^\dagger_{X_{\gamma} \ell m_{\gamma} (p) \omega}\hat a_{X'_{\gamma} \ell' m'_{\gamma} (p) \omega'}\rangle = f^{\gamma}_{X_{\gamma} X'_{\gamma} \ell (p)}(\omega)\,2\pi \delta(\omega - \omega')\delta_{\ell \ell'}\delta_{m_{\gamma},m'_{\gamma}},
\end{equation}
and similarly for the electron and positron. Note that this expectation value is diagonal in $\ell$, $m_\gamma$, and $p$, and independent of $m_\gamma$, due to spherical symmetry and parity. For free fields, it is diagonal in $\omega$ due to time translation invariance. (We will address the subtleties with interacting fields below.) The symmetries of the problem allow the ``in'' and ``up'' modes to become correlated, so $f^\gamma_{X_{\gamma} X'_{\gamma} \ell(p)}(\omega)$ is a $2\times 2$ Hermitian matrix for each $\ell$, $p$, and $\omega$.

To take into account stimulated emission (for photons) and Pauli blocking (for fermions), we need to define the enhanced occupation function (for photons) or unoccupation functions (for fermions):
\begin{eqnarray}
\label{eq: phase_density_photonII}
\langle \hat a_{X_{\gamma \ell m_{\gamma}} \omega (p)} \hat a^\dagger_{X'_{\gamma \ell' m'_{\gamma}} \omega' (p)} \rangle &=& 2\pi g^{\gamma}_{X_{\gamma}X'_{\gamma} \ell  (p)}(\omega)\delta(\omega-\omega')\delta_{\ell \ell'} \delta_{m_{\gamma}m'_{\gamma}} ~~~{\rm where}
\nonumber \\
g^{\gamma}_{X_{\gamma}X'_{\gamma} \ell  (p)}(\omega) &\equiv& \delta_{X_{\gamma}X'_{\gamma}} + f^{\gamma \ast}_{X_{\gamma}X'_{\gamma} \ell  (p)}(\omega).
\end{eqnarray}
The definition for $g$ has a $+$ sign for bosonic operators and a $-$ sign for fermionic operators due to the commutation/anticommutation relations for either operator type:
\begin{equation}
g^{e^-}_{XX'k}(h) \equiv \delta_{XX'} - f^{e^- \ast}_{XX'k}(h).
\end{equation}

For the non-interacting fields, the phase space densities are given by the usual formulae:
\begin{equation}
\label{eq: phase_densities}
f^\gamma_{{\rm up,up,} \ell (p)}(\omega) = \frac1{e^{8\pi M\omega}-1}, ~~~
f^{e^-\,{\rm or}\,e^+}_{{\rm up,up,} k}(h) = \frac1{e^{8\pi Mh}+1},
\end{equation}
and 0 for the other (in,in; in,up; or up,in) components. Here $1/(8\pi M)$ is the Hawking temperature of the black hole. The outgoing phase space density, from Eq.~(\ref{eq: out_out_intensity}), is
\begin{equation}
f^\gamma_{{\rm out,out,} \ell (p)}(\omega) = \frac{|T_{1,\ell,\omega}|^2}{e^{8\pi M\omega}-1},
\end{equation}
and then the spectrum of outgoing photons in non-interacting theory \cite{1976PhRvD..13..198P} is
\begin{equation}
\frac{dN_\gamma^{(0)}}{d\omega\,dt} = \frac1{2\pi} \sum_{\ell m_\gamma p} f^\gamma_{{\rm out,out,} \ell (p)}(\omega)
= \frac1\pi \sum_{\ell=1}^\infty (2\ell+1) \frac{|T_{1,\ell,\omega}|^2}{e^{8\pi M\omega}-1}.
\label{eq:non-int-out}
\end{equation}
Note that here we have taken the spectrum per unit $\omega$, so $\int d\omega/(2\pi) \rightarrow 1/(2\pi)$, and we have performed the trivial sums over $m_\gamma$ and $p$ to get a factor of $2(2\ell+1)$. The correction to the emitted radiation due to interactions is (by comparison to Eq.~\ref{eq:HEMo}, and recognizing that photon number differs from energy by a factor of $\omega$):
\begin{equation}
\frac{dN_\gamma^{(1)}}{d\omega\,dt} = \frac1{2\pi} \sum_{\ell m_\gamma p} \frac{d}{dt} \langle \hat a^\dagger_{{\rm out},\ell,m_\gamma,\omega(p)} \hat a_{{\rm out},\ell,m_\gamma,\omega(p)} \rangle.
\label{eq:ddt-out}
\end{equation}
Our principal aim in this series of papers is to evaluate Eq.~(\ref{eq:ddt-out}).

\subsection{Evolution due to the interaction Hamiltonian}

We now compute the evolution of the photon density matrix, following the procedure in \citet{2004ASSL..307.....L}. We will simplify our task by working only to order $e^2$.

We will define interaction picture operators via $\hat{O}_{\rm I} = e^{\rm i H_{\rm Dirac + \gamma}t} \hat{O} e^{-\rm i H_{\rm Dirac + \gamma}t}$, where $H_{\rm Dirac + \gamma} = H_{\rm Dirac} + H_{\rm \gamma}$ is the free-field Hamiltonian for the non-interacting electron and photon fields. We will need the usual commutation relations for the photon and electron operators:
\begin{equation}
\label{eq: H_commutators}
\begin{array}{lll}
~[H_{\gamma},\hat a_{X_{\gamma},\ell,m_{\gamma},\omega,(p)}] = -\omega \hat a_{X_{\gamma},\ell,m_{\gamma},\omega,(p)}, & & [H_{\gamma},\hat a^\dagger_{X_{\gamma},\ell,m_{\gamma},\omega,(p)}] = \omega \hat a^\dagger_{X_{\gamma},\ell,m_{\gamma},\omega,(p)},
\\
~[H_{\rm Dirac},\hat b_{Xkmh} ] = -h \hat b_{Xkmh} ,  & & [H_{\rm Dirac},\hat b^\dagger_{Xkmh} ] = h \hat b^\dagger_{Xkmh},
\\
~[H_{\rm Dirac}, \hat d_{Xkmh}] = -h \hat d_{Xkmh}, &
~~{\rm and}~~~~~~~
& [H_{\rm Dirac}, \hat d^\dagger_{Xkmh}] = h \hat d^\dagger_{Xkmh}.
\end{array}
\end{equation}

Now we can write the interaction picture operators:
\begin{equation}
H_{\rm int,I}(t) \equiv e^{\rm i H_{\rm Dirac + \gamma}t} H_{\rm int} e^{-\rm i H_{\rm Dirac + \gamma}t} = B(t) + B^\dagger (t),
\label{eq: B_int}
\end{equation}
where
\begin{equation}
B(t) = e \sum_{X_\gamma\ell m_\gamma p} \int \frac{d\omega}{2\pi}  \hat a_{X_{\gamma},\ell,m_{\gamma},\omega,(p)}e^{-i\omega t} Q^{(p)}_{X_\gamma \ell m_\gamma,\omega}.
\end{equation}
The factors with the photon operators removed from them are denoted $Q(t)$ such that we have the form $H_{\rm int} \propto \hat a Q(t)$. This form explicitly separates the electromagnetic operators from electron/positron operators:
\begin{eqnarray}
Q^{(p)}_{X_\gamma \ell m_\gamma,\omega} &=& 
\int \frac{dh\,dh'}{(2\pi)^2}
\sum_{XkmX'k'm'}
\Bigl[ \hat b^\dagger_{Xkmh} \hat d^\dagger_{X'k'm'h'} e^{i(h+h')t} I^{++}_{Xkm,X'k'm',X_\gamma \ell m_\gamma (p)}  
\nonumber \\
&& +
\hat d_{Xkmh} \hat b_{X'k'm'h'} e^{-i(h+h')t} I^{--}_{X-km,X'k'm',X_\gamma \ell m_\gamma (p)}  
+
\hat b^\dagger_{Xkmh} \hat b_{X'k'm'h'} e^{i(h-h')t} I^{+-}_{Xkm,X'k'm',X_\gamma \ell m_\gamma (p)} 
\nonumber \\ &&
+ \hat d_{Xkmh} \hat d^\dagger_{X'k'm'h'} e^{-i(h-h')t} I^{-+}_{Xkm,X'k'm',X_\gamma \ell m_\gamma (p)}
\Bigr].
\end{eqnarray}
Here we used Eq.~(\ref{eq: H_commutators}) to rewrite terms such as $e^{iH_{\rm Dirac}t} \hat b^\dagger_{s} \hat d^\dagger_{s'} e^{-iH_{\rm Dirac}t} = e^{i(h+h')t}\hat b^\dagger_{s} \hat d^\dagger_{s'}$.

Now that we have the interaction Hamilitonian in the interaction picture, Eq.~(\ref{eq: B_int}), we now follow the procedure that led to Eq.~(6.74) of \citet{2004ASSL..307.....L} to arrive at
\begin{equation}
\label{eq: operator_evolution}
\frac{d}{dt}\langle \hat a^\dagger_{\ell m_{\gamma} \omega (p) X_{\gamma}} \hat a_{\ell m_{\gamma} \omega (p) X'_{\gamma}} \rangle = -{\rm Tr\,}\Bigl(\int^t_0 dt' [[\hat a^\dagger_{X_{\gamma} \ell m_{\gamma} \omega (p)} \hat a_{X'_{\gamma} \ell m_{\gamma} \omega (p)},B(t)],B^\dagger (t')] \rho_{\rm I}(t') \Bigr) + ({\rm c.c.,~} X_{\gamma} \Leftrightarrow X_{\gamma}'),
\end{equation}
where $\rho_{\rm I}(t')$ is the interaction picture density matrix at time $t'$. Since the initial conditions for the photon (Hawking radiation) are an incoherent state, the same removal of terms that do not have both a $B$ and a $B^\dagger$ (Eqs.~6-45--6-47 of \citet{2004ASSL..307.....L}) is valid here. We may now write the pieces of this expression. We start with the inner commutator,
\begin{equation}
\left[\hat a^\dagger_{X_{\gamma} \ell m_{\gamma} \omega (p) } \hat a_{X'_{\gamma} \ell m_{\gamma} \omega (p)},B(t)\right] = -e \sum_p (\hat a_{X'_{\gamma} \ell m_{\gamma} \omega (p) } e^{-i\omega t}Q^{(p)}_{X_{\gamma} \ell m_{\gamma} \omega}) ;
\end{equation}
then proceed to the outer commutator,
\begin{eqnarray}
\left[\left[\hat a^\dagger_{\ell m_{\gamma} \omega (p) X_{\gamma}} \hat a_{\ell m_{\gamma} \omega (p) X'_{\gamma}},B(t)\right],B^\dagger (t')\right] &=& e^2 
\int \frac{d\omega''}{2\pi} \sum_{X''_\gamma \ell''m''_\gamma p''}
\Bigl[ (\hat a_{X'_{\gamma} \ell m_{\gamma} (p) \omega}\hat a^\dagger_{X''_{\gamma} \ell'' m''_{\gamma} (p'') \omega''} Q^{(p)}_{X_{\gamma} \ell m_{\gamma} \omega} Q^{(p'')\dagger}_{X''_{\gamma} \ell'' m''_{\gamma} \omega''} 
\nonumber \\ && -
\hat a^\dagger_{X''_{\gamma} \ell'' m''_{\gamma} (p'') \omega''}\hat a_{X'_{\gamma} \ell m_{\gamma} (p) \omega} Q^{(p'') \dagger}_{X''_{\gamma} \ell'' m''_{\gamma} \omega''} Q^{(p)}_{X_{\gamma} \ell m_{\gamma} \omega}) e^{-i(\omega t - \omega'' t')} \Bigr];
\label{eq:statavgint}
\end{eqnarray}
and then proceed to the statistical average:
\begin{eqnarray}
&& \!\!\!\!\!\!\!\!\!\!\!\!\!\!\!\!
{\rm Tr}\Bigl( \left[\left[\hat a^\dagger_{\ell m_{\gamma} \omega (p) X_{\gamma}} \hat a_{\ell m_{\gamma} \omega (p) X'_{\gamma}},B(t)\right],B^\dagger (t')\right] \rho_{\rm I}(t') \Bigr) 
\nonumber \\ &=&
e^2 (\langle Q^{(p)}_{X_{\gamma} \ell m_{\gamma} \omega} Q^{(p)\dagger}_{X'_{\gamma} \ell m_{\gamma} \omega}\rangle + \sum_{X''_{\gamma}} f^{\gamma *}_{X'_{\gamma}X''_{\gamma} \ell m_{\gamma}}(\omega) \langle [ Q^{(p)}_{X_{\gamma} \ell m_{\gamma} \omega}, Q^{(p)\dagger}_{X''_{\gamma} \ell m_{\gamma} \omega} ]\rangle) e^{-i \omega(t - t')}.
\label{eq: stat_avg}
\end{eqnarray}
We can further simplify this expression using the occupation functions (phase space densities) as given in Eqs.~(\ref{eq: phase_density_photonI}) and (\ref{eq: phase_density_photonII}). This, combined with the expectation values of the different fermionic operators (see Appendix~\ref{app: four_fermion_op} for details of how to simplify the expectation values of the 4-fermion operators), gives
\begin{eqnarray}
&& \!\!\!\!\!\!\!\!\!\!\!\!\!\!\!\!
\frac{d}{dt}\langle \hat a^\dagger_{\ell m_{\gamma} \omega (p) X_{\gamma}} \hat a_{\ell m_{\gamma} \omega (p) X'_{\gamma}} \rangle 
\nonumber \\
&=& e^2 \int \frac{dh\,dh'}{(2\pi)^2} \sum_{XkmX'k'm'X''X'''} \pi
\nonumber \\ && \times
\Bigl( f^{e^-}_{XX'''k}(h) f^{e^+}_{X'X''k'}(h') I^{++}_{X km,X' k'm',X_{\gamma} \ell m_\gamma (p)}(h,h',\omega) I^{++\ast}_{X''' km, X'' k'm',X'_{\gamma} \ell m_\gamma (p)}(h,h',\omega) \Phi(\omega - h-h') 
\nonumber \\ && ~~~~+
g^{e^+}_{XX'''k}(h)g^{e^-}_{X'X''k'}(h')I^{--}_{Xkm, X' k'm',X_{\gamma} \ell m_\gamma (p)}(h,h',\omega) I^{--\ast}_{X''' km, X'' k'm',X'_{\gamma} \ell m_\gamma (p)}(h,h',\omega) \Phi(\omega + h + h')
\nonumber \\ && ~~~~+
f^{e^-}_{XX'''k}(h) g^{e^-}_{X'X''k'}(h')
I^{+-}_{X km, X' k'm',X_{\gamma} \ell m_\gamma (p)}(h,h',\omega) I^{+-\ast}_{X''' km, X'' k'm', X'_{\gamma} \ell m_\gamma (p)}(h,h',\omega) \Phi(\omega + h'-h) 
\nonumber \\ && ~~~~+
g^{e^+}_{XX'''k}(h)f^{e^+}_{X'X''k'}(h') I^{-+}_{X km, X' k'm',X_{\gamma} \ell m_\gamma (p)}(h,h',\omega) I^{-+\ast}_{X''' km, X'' k'm',X'_{\gamma} \ell m_\gamma (p)}(h,h',\omega) \Phi(h-h' + \omega) 
\nonumber \\ && +
\sum_{X''_{\gamma}} f^{\gamma \ast}_{X'_{\gamma}X''_{\gamma} \ell (p)}(\omega) \Bigl\{[f^{e^-}_{XX'''k}(h) f^{e^+}_{X'X''k'}(h') - g^{e^+}_{X''X'k'}(h')g^{e^-}_{X'''Xk}(h)]
\nonumber \\ && ~~~~~~~~\times
I^{++}_{X km, X' k'm',X_{\gamma} \ell m_\gamma (p)}(h,h',\omega)I^{++\ast}_{X''' km, X'' k'm', X''_{\gamma} \ell m_\gamma (p)}(h,h',\omega) \Phi(\omega - h-h') 
\nonumber \\ && ~~~~+
[g^{e^+}_{XX'''k}(h)g^{e^-}_{X'X''k'}(h') - f^{e^-}_{X''X'k'}(h')f^{e^+}_{X'''Xk}(h)]
\nonumber \\ && ~~~~~~~~\times
I^{--}_{Xkm,X' k'm',X_{\gamma} \ell m_\gamma (p)}(h,h',\omega) I^{--\ast}_{X''' km,X'' k'm',X''_{\gamma} \ell m_\gamma (p)}(h,h',\omega)  \Phi(\omega + h + h') 
\nonumber \\ && ~~~~+
[f^{e^-}_{XX'''k}(h)g^{e^-}_{X'X''k'}(h') - g^{e^-}_{X'''Xk}(h)f^{e^-}_{X''X'k'}(h')]
\nonumber \\ && ~~~~~~~~\times
I^{+-}_{Xkm,X'k'm',X_{\gamma}\ell m_\gamma(p)}(h,h',\omega) I^{+-\ast}_{X'''km,X''k'm',X''_{\gamma}\ell m_\gamma (p)}(h,h',\omega) \Phi(\omega + h' - h) 
\nonumber\\ && ~~~~+
[g^{e^+}_{XX'''k}(h)f^{e^+}_{X'X''k'}(h') - f^{e^+}_{X'''Xk}(h) g^{e^+}_{X''X'k'}(h')]
\nonumber \\ && ~~~~~~~~\times
I^{-+}_{Xkm,X'k'm',X_{\gamma}\ell m_\gamma (p)}(h,h',\omega) I^{-+\ast}_{X'''km,X''k'm',X''_{\gamma}\ell m_\gamma(p)}(h,h',\omega) \Phi(h-h' + \omega)\Bigr\}\Bigr)
\nonumber\\ &&
+({\rm c.c.},~ X_{\gamma} \Leftrightarrow X_{\gamma}'),
\label{eq: nested_1}
\end{eqnarray}
In arriving at Eq.~(\ref{eq: nested_1}), we used the simplification for coupling integrals with the same $m$-index for different spinor state, e.g. $I^{\tau\tau'}_{X km,X' km,X_{\gamma} \ell m_\gamma (p)}(h,h,\omega)$, the sum over $m$ will simplify this to zero by symmetry of the 3$j$ symbols:
\begin{equation}
\sum_{m=-j}^j (-1)^{m-1/2}\tj{j}{j}{\ell}{-m}{m}{m_\gamma} = 0
~~~{\rm for}~~~ \ell\ge 1.
\end{equation}
This eliminates several of the possible contractions that arise from simplification of the 4-fermion operators. Physically, this corresponds to the suppression of transitions with fermions of angular momentum $j,m$ of state $X$ transitioning to the state $X'$ with the \textit{same} $j,m$ during the emission or absorption a photon of angular momentum $\ell \ge 1$ since this transition would violate conservation of angular momentum (or stated in terms of Feynman diagrams, tadpole diagrams are disallowed). So only terms with coupling integrals over different $k,m$-indices will contribute to the evolution of the photon density matrix.

We have also substituted $\Phi(\Omega)$ for the time integral defined in Eq.~(6.57) of \citet{2004ASSL..307.....L}. The $\Phi(\Omega)$ is defined as the improper integral of a complex exponential:
\begin{equation}
\label{eq: Phi_Omega}
\Phi(\Omega) = \frac1\pi \lim_{t\rightarrow \infty} \int_{0}^{t}e^{i\Omega(t-t')}\,dt' = \delta(\Omega) + \frac{i}{\pi} {\rm P}\frac{1}{\Omega},
\end{equation}
where ${\rm P}(1/\Omega)$ is the principal part in the distribution sense (improper integral sense). Eq.~(\ref{eq: Phi_Omega}) has two parts that contribute to different types of interactions: a \textit{dissipative} part ($\delta(\Omega)$) and a \textit{conservative} part (${\rm P}(1/\Omega)$). The dissipative part is related to processes involving the emission and absorption of photons from the different spinor fields (pair production, pair annihilation, and bound-bound, bound-free, and free-free transitions of the electrons and positrons). The conservative part (which includes plasma frequency and vacuum polarization effects) does not change the total number or energy of photons, but will affect the barrier transmission probability and hence the emitted Hawking radiation. We can therefore break Eq.~(\ref{eq: operator_evolution}) into dissipative and conservative pieces: 
\begin{equation}
\frac{d}{dt}\langle \hat a^\dagger_{X_{\gamma} \beta (p)} \hat a_{X'_{\gamma}\beta (p)} \rangle = \frac{d}{dt}\langle \hat a^\dagger_{X_{\gamma} \beta (p)} \hat a_{X'_{\gamma}\beta (p)} \rangle_{\rm diss} + \frac{d}{dt}\langle \hat a^\dagger_{X_{\gamma} \beta (p)} \hat a_{X'_{\gamma}\beta (p)} \rangle_{\rm cons},
\label{eq:diss-cons}
\end{equation}
where the dissipative terms come from the $ \delta(\Omega)$ and the conservative terms come from the $ {\rm P}(1/\Omega)$ in Eq.~(\ref{eq: Phi_Omega}).

\subsection{Dissipative part of the photon density evolution}

The remainder of this calculation will deal with the dissipative part of Eq.~(\ref{eq: Phi_Omega}), and a later paper in this series will handle the contribution due to the conservative part.

If we consider only the dissipative part, we can further simplify Eq.~(\ref{eq: nested_1}) by the following sequence of simplifications:
\begin{enumerate}
    \item We collapse the energy integral $\int dh'$ using the $\delta$-function.
    \item We drop the $I^{--}$ terms since the $\delta$-function condition $h' = -(\omega + h)$ does not contribute to the integrals over positive energies (physically, we cannot have an on-shell photon, electron, and positron appear out of the vacuum or disappear into it).
    \item We simplify the $\sum_{mm'}$ over products of the $3j$ symbols in Eq.~(\ref{eq:double-bracket}) using Eq.~(3.7.8) of \citet{1960amqm.book.....E}, giving an overall factor of $\Delta(j,j',\ell)/(2\ell + 1)$.
    \item We combine the terms with coupling integrals $\llbracket I^{+-}_{Xk,X'k',X_{\gamma}\ell (p)}\rrbracket \llbracket I^{+-*}_{Xk,X'k',X'_{\gamma}\ell (p)}\rrbracket$ (corresponding to the process $e^-+\gamma\leftrightarrow e^-$) and $\llbracket I^{-+}_{Xk,X'k',X_{\gamma}\ell (p)}\rrbracket \llbracket I^{-+*}_{Xk,X'k',X'_{\gamma}\ell (p)} \rrbracket$ (corresponding to $e^++\gamma\leftrightarrow e^+$) since they are the same under charge conjugation ($\hat{C}$ symmetry). Physically, emission and absorption from electron levels and positron levels give the same contribution, so we have a factor of 2. The $\hat C$ symmetry is explored in Appendix~\ref{app:C}, leading specifically to the result that allows us to combine the terms, Eq.~(\ref{eq: I_C_symmetry}). 
    \item We work in the \{in,up\} basis, where the phase space densities $f_{X'X''}$ and $g_{X'X''}$ are diagonal.
\end{enumerate}
This gives us:
\begin{eqnarray}
&& \!\!\!\!\!\!\!\!\!\!\!\!
\frac{d}{dt}\langle \hat a^\dagger_{X_{\gamma} \ell m_{\gamma} \omega (p)} \hat a_{X'_{\gamma}\ell m_{\gamma} \omega (p)} \rangle_{\rm diss}
\nonumber \\ &=&
\frac{e^2}{2(2\ell+1)} \int \frac{dh}{2 \pi} \sum_{X,X' \in \left\{{\rm in,up} \right\}}
\sum_{kk'}
\Delta(j,j',\ell) \delta_{ss'(-1)^{k+k'+\ell},(-1)^p}
\nonumber \\ && \times
\Bigl( f^{e^-}_{XX}(h) f^{e^+}_{X'X'}(h') \llbracket I^{++}_{X k,X' k',X_{\gamma} \ell (p)}(h,h',\omega) \rrbracket \llbracket I^{++*}_{X k, X' k',X'_{\gamma} \ell (p)}(h,h',\omega) \rrbracket |_{h' = \omega - h}
\nonumber \\ && +
2f^{e^+}_{X'X'}(h') g^{e^+}_{XX}(h) \llbracket I^{-+}_{X k, X' k',X_{\gamma} \ell (p)}(h,h',\omega) \rrbracket \llbracket I^{-+*}_{X k, X'k', X'_{\gamma} \ell (p)}(h,h',\omega) \rrbracket |_{h' = h + \omega} 
\nonumber \\ && +
f^{\gamma *}_{X'_{\gamma}X'_{\gamma} \ell}(\omega) \Bigl\{ (f^{e^-}_{XX}(h) f^{e^+}_{X'X'}(h') - g^{e^+}_{X'X'}(h') g^{e^-}_{XX}(h)) 
\nonumber \\ &&~~~~ \times
\llbracket I^{++}_{X k, X' k',X_{\gamma} \ell (p)}(h,h',\omega) \rrbracket \llbracket I^{++*}_{X k, X' k', X'_{\gamma} \ell (p)}(h,h',\omega) \rrbracket |_{h' = \omega - h} 
\nonumber \\ && +
2(f^{e^+}_{X'X'}(h') g^{e^+}_{XX}(h) - g^{e^+}_{X'X'}(h')f^{e^+}_{XX}(h)) \llbracket I^{-+}_{X k,X'k',X_{\gamma}\ell(p)}(h,h',\omega) \rrbracket \llbracket I^{-+*}_{Xk,X'k',X'_{\gamma}\ell(p)}(h,h',\omega) \rrbracket |_{h' = h + \omega} \Bigr\}\Bigr)
\nonumber\\ &&
+~ ({\rm c.c.},~ X_{\gamma} \Leftrightarrow X_{\gamma}'),
\label{eq: evolve_diss}
\end{eqnarray}
where we define one of the energy index for the spinor fields $h'$ to depend on some combination of the other spinor field energy $h$ and photon energy $\omega$, and $\Delta(j,j',\ell) \delta_{ss'(-1)^{k+k'+\ell},(-1)^p}$ encodes the angular momentum and parity selection rules. We can further simplify Eq.~(\ref{eq: evolve_diss}) by using the phase space densities for the electrons and positrons in Eq.~(\ref{eq: phase_densities}) to give
\begin{eqnarray}
&& \!\!\!\!\!\!\!\!\!\!\!\!
\frac{d}{dt}\langle \hat a^\dagger_{X_{\gamma} \ell m_{\gamma} \omega (p)} \hat a_{X'_{\gamma}\ell m_{\gamma} \omega (p)} \rangle_{\rm diss}
\nonumber \\ &=&
\frac{e^2}{2(2\ell+1)} \int \frac{dh}{2 \pi}
\sum_{kk'}
\Delta(j,j',\ell) \delta_{ss'(-1)^{k+k'+\ell},(-1)^p}
\nonumber \\ && \times
\Bigl[ -f^{\gamma \ast}_{X'_\gamma X'_\gamma \ell} (\omega) \llbracket I^{++}_{{\rm in}~k, {\rm in}~k',X_{\gamma} \ell (p)}(h,\omega - h,\omega) \rrbracket \llbracket I^{++*}_{{\rm in}~k, {\rm in}~k',X'_{\gamma} \ell (p)}(h,\omega - h,\omega) \rrbracket
\nonumber \\ && -
f^{\gamma \ast}_{X'_\gamma X'_\gamma \ell} (\omega) \frac{e^{8\pi M (\omega - h)}}{e^{8 \pi M (\omega - h)} + 1} \llbracket I^{++}_{{\rm in}~k, {\rm up}~k',X_{\gamma} \ell (p)}(h,\omega - h,\omega) \rrbracket \llbracket I^{++*}_{{\rm in}~k, {\rm up}~k',X'_{\gamma} \ell (p)}(h,\omega - h,\omega) \rrbracket 
\nonumber \\ && + (1+f^{\gamma \ast}_{X'_\gamma X'_\gamma \ell} (\omega))
\frac{2}{e^{8 \pi M (\omega + h)} + 1} \llbracket I^{-+}_{{\rm in} ~k, {\rm up}~k',X_{\gamma} \ell (p)}(h,\omega + h,\omega) \rrbracket \llbracket I^{-+*}_{{\rm in}~k, {\rm up}~k', X'_{\gamma} \ell (p)}(h,\omega + h,\omega) \rrbracket 
\nonumber \\ && -
 f^{\gamma \ast}_{X'_\gamma X'_\gamma \ell} (\omega) \Bigl( \frac{e^{8\pi M h}}{e^{8 \pi M h} + 1} \llbracket I^{++}_{{\rm up}~k, {\rm in}~k',X_{\gamma} \ell (p)}(h,\omega - h,\omega) \rrbracket \llbracket I^{++*}_{{\rm up}~k, {\rm in}~k',X'_{\gamma} \ell (p)}(h,\omega - h,\omega) \rrbracket
\nonumber \\ && ~~~~+
\frac{2}{e^{8 \pi M h} + 1} \llbracket I^{-+}_{{\rm up} ~k, {\rm in}~k',X_{\gamma} \ell (p)}(h,\omega + h,\omega) \rrbracket \llbracket I^{-+*}_{{\rm up}~k, {\rm in}~k', X'_{\gamma} \ell (p)}(h,\omega + h,\omega) \rrbracket \Bigr) 
\nonumber \\ && +
[1 + (1 - e^{8 \pi M \omega})f^{\gamma \ast}_{X'_\gamma X'_\gamma \ell}(\omega)] 
\nonumber \\ && ~~~~\times
\Bigl( \frac{1}{(e^{8 \pi M h} + 1)(e^{8 \pi M (\omega - h)} + 1)} \llbracket I^{++}_{{\rm up}~k, {\rm up}~k',X_{\gamma} \ell (p)}(h,\omega - h,\omega) \rrbracket \llbracket I^{++*}_{{\rm up}~k, {\rm up}~k',X'_{\gamma} \ell (p)}(h,\omega - h,\omega) \rrbracket
 \nonumber \\ && ~~~~ + 
 \frac{2 e^{8 \pi M h}}{(e^{8 \pi M h} + 1)(e^{8 \pi M (\omega + h)} + 1)} \llbracket I^{-+}_{{\rm up} ~k, {\rm up}~k',X_{\gamma} \ell (p)}(h,\omega + h,\omega) \rrbracket \llbracket I^{-+*}_{{\rm up}~k, {\rm up}~k', X'_{\gamma} \ell (p)}(h,\omega + h,\omega) \rrbracket \Bigr) \Bigr]
\nonumber\\ &&
+~ ({\rm c.c.},~ X_{\gamma} \Leftrightarrow X_{\gamma}'),
\label{eq: evolve_diss_simplify}
\end{eqnarray}
where we grouped terms of the same $X,X'$ type together with the appropriate prefactors of photon phase space density, $f^{\gamma}(\omega)$. The $(X,X')=({\rm in,in})$ terms are grouped first; then the (in,up) terms; then the (up,in) terms; and lastly the (up,up) terms. We can see after simplifications, there are three terms that contribute to the spontaneous emission (that exist even when $f^{\gamma \ast}(\omega)=0$) coming from the set of scattering solutions $(X,X') = ({\rm in,up})$ and $({\rm up,up})$. These spontaneous emission terms correspond to:
\begin{list}{$\bullet$}{}
    \item the production of photons from electron-positron annihilation, $e^++e^-\rightarrow\gamma$, described by the $I^{++}$ term with fermion indices (up,up), and where the energies of the electron and positron sum to $\omega$; and
    \item the braking radiation (inner bremsstrahlung) of the electron or positron, $e^\mp \rightarrow e^\mp + \gamma$, and where the energies of the two fermion states have a difference of $\omega$. This is associated with the $I^{-+}$ terms with fermion indices (up,up) and (in,up). The factor of 2 comes from inclusion of both signs of charge.
\end{list}
There are also absorption terms, which contain a $-f^{\gamma\ast}_{X'_\gamma X'_\gamma}(\omega)$, as well as stimulated emission terms.
For the specific case where both $X_\gamma = X'_\gamma = {\rm up}$, the $(X,X')=({\rm up,up})$ terms completely cancel, since then $1 + (1 - e^{8 \pi M \omega})f^{\gamma \ast}_{X'_\gamma X'_\gamma \ell}(\omega)=0$. Physically, this is a result of thermodynamic equilibrium: the Hawking radiation coming up from the horizon has the same temperature $T_{\rm H}=1/(8\pi M)$ for all particle species, so the net rate of change of occupation numbers due to interactions of the ``up'' particles is zero. 

The ${\cal O}(\alpha)$ correction to the emitted radiation spectrum, Eq.~(\ref{eq:ddt-out}), is related to this phase space density evolution via Eq.~(\ref{eq: out_out_intensity}). We separate Eq.~(\ref{eq: evolve_diss_simplify}) into the ``direct'' terms (written explicitly) and the ``c.c.s.'' (complex conjugate with swap) terms. We see that
\begin{eqnarray}
\frac{d}{dt}\langle \hat a^\dagger_{{\rm out}, \ell m_{\gamma} \omega (p)} \hat a_{{\rm out},\ell m_{\gamma} \omega (p)} \rangle_{\rm diss}
 &=& 
2 \bigl\{
|R_{1,\ell\omega}|^2 [{\rm direct\,term\,in,in}]
+|T_{1,\ell\omega}|^2 [{\rm direct\,term\,up,up}]
\nonumber \\ && ~~~~
+\Re R_{1,\ell\omega}^\ast T_{1,\ell\omega}
([{\rm direct\,term\,in,up}] + [{\rm direct\,term\,up,in}]^\ast)
\bigr\}.
\label{eq:temp-out-out}
\end{eqnarray}

We are now able to now write the dissipative correction to the emitted radiation to order $e^2$:
\begin{eqnarray}
&& \!\!\!\!\!\!\!\!\!\!\!\!
\left.\frac{dN_\gamma^{(1)}}{dt\,d\omega}\right|_{\rm diss} = \frac{1}{2 \pi} \sum_{\ell m_\gamma p} \frac{d}{dt}\langle \hat a^\dagger_{{\rm out},\ell m_{\gamma} \omega (p)} \hat a_{{\rm out},\ell m_{\gamma} \omega (p)} \rangle_{\rm diss}
\nonumber \\ &=&
\frac{e^2}{2 \pi} \sum_{\ell = 0}^\infty \sum_p \int \frac{dh}{2 \pi}
\sum_{kk'}
\Delta(j,j',\ell) \delta_{ss'(-1)^{k+k'+\ell},(-1)^p}
\nonumber \\ && \times
\Bigl[ |R_{1,\ell,\omega}|^2 \Bigl\{ \frac{2}{e^{8 \pi M (\omega + h)} + 1} | \llbracket I^{-+}_{{\rm in},k,{\rm up},k',{\rm in},\ell,(p)}(h,\omega + h,\omega) \rrbracket|^2  
\nonumber \\ && + 
\frac{1}{(e^{8 \pi M h} + 1)(e^{8 \pi M (\omega - h)} + 1)}| \llbracket I^{++}_{{\rm up},k,{\rm up},k',{\rm in},\ell,(p)}(h,\omega - h,\omega) \rrbracket|^2 
\nonumber \\ && + 
\frac{2e^{8 \pi M h}}{(e^{8 \pi M h} + 1)(e^{8 \pi M (\omega + h)} + 1)} | \llbracket I^{-+}_{{\rm up},k,{\rm up},k',{\rm in},\ell,(p)}(h,\omega + h,\omega) \rrbracket|^2 \Bigr\} 
\nonumber \\ && + 
\frac{|T_{1,\ell,\omega}|^2}{e^{8 \pi M \omega} - 1} \Bigl\{-| \llbracket I^{++}_{{\rm in},k,{\rm in},k',{\rm up},\ell,(p)}(h,\omega - h,\omega) \rrbracket|^2 +
\frac{2e^{8 \pi M \omega}}{e^{8 \pi M (\omega + h)} + 1} | \llbracket I^{-+}_{{\rm in},k,{\rm up},k',{\rm up},\ell,(p)}(h,\omega + h,\omega) \rrbracket|^2 
\nonumber \\ && -
\frac{2}{e^{8 \pi M h} + 1}\Bigl( e^{8 \pi M h}| \llbracket I^{++}_{{\rm up},k,{\rm in},k',{\rm up},\ell,(p)}(h,\omega - h,\omega) \rrbracket|^2 +
| \llbracket I^{-+}_{{\rm up},k,{\rm in},k',{\rm up},\ell,(p)}(h,\omega + h,\omega) \rrbracket|^2\Bigr)\Bigr\} +
\nonumber \\ && + 
\Re T^\ast_{1,\ell,\omega}R_{1,\ell,\omega}\Bigl[ \frac{2(2e^{8 \pi M \omega} - 1)}{(e^{8 \pi M (\omega + h)} + 1)(e^{8 \pi M \omega} - 1)} \llbracket I^{-+}_{{\rm in},k,{\rm up},k',{\rm up},\ell,(p)}(h,\omega + h,\omega) \rrbracket \llbracket I^{-+*}_{{\rm in},k,{\rm up},k',{\rm in},\ell,(p)}(h,\omega + h,\omega) \rrbracket 
\nonumber \\ && + 
\frac{1}{(e^{8 \pi M h} + 1)(e^{8 \pi M (\omega - h)} + 1)} \llbracket I^{++}_{{\rm up},k,{\rm up},k',{\rm up},\ell,(p)}(h,\omega - h,\omega) \rrbracket \llbracket I^{++ \ast}_{{\rm up},k,{\rm up},k',{\rm in},\ell,(p)}(h,\omega - h,\omega) \rrbracket  
\nonumber \\ && + 
\frac{2e^{8 \pi M h}}{(e^{8 \pi M h} + 1)(e^{8 \pi M (\omega + h)} + 1)} \llbracket I^{-+}_{{\rm up},k,{\rm up},k',{\rm up},\ell,(p)}(h,\omega + h,\omega) \rrbracket \llbracket I^{-+\ast}_{{\rm up},k,{\rm up},k',{\rm in},\ell,(p)}(h,\omega + h,\omega) \rrbracket 
\nonumber \\ && - 
\frac{1}{e^{8 \pi M \omega} - 1} \Bigl( \llbracket I^{++}_{{\rm in},k,{\rm in},k',{\rm up},\ell,(p)}(h,\omega - h,\omega) \rrbracket \llbracket I^{++ \ast}_{{\rm in},k,{\rm in},k',{\rm in},\ell,(p)}(h,\omega - h,\omega) \rrbracket 
\nonumber \\ && +
\frac{2e^{8 \pi M h}}{e^{8 \pi M h} + 1} \llbracket I^{++}_{{\rm up},k,{\rm in},k',{\rm up},\ell,(p)}(h,\omega - h,\omega) \rrbracket \llbracket I^{++ \ast}_{{\rm up},k,{\rm in},k',{\rm in},\ell,(p)}(h,\omega - h,\omega) \rrbracket 
\nonumber \\ && +
\frac{2}{e^{8 \pi M h} + 1} \llbracket I^{-+}_{{\rm up},k,{\rm in},k',{\rm up},\ell,(p)}(h,\omega + h,\omega) \rrbracket \llbracket I^{-+ \ast}_{{\rm up},k,{\rm in},k',{\rm in},\ell,(p)}(h,\omega + h,\omega) \rrbracket \Bigr) \Bigr]  \Bigr],
\label{eq: emitted_correction}
\end{eqnarray}
where the energy integral for each term is implied to be over the legal range for all arguments in the $I$-integrals (energies positive, and $>\mu$ in the case of the fermion ``in'' modes).
In this expression, we have made several algebraic simplifications:
\begin{list}{$\bullet$}{}
\item The $\sum_{m_\gamma}$ cancels the factor of $1/(2\ell + 1)$ since there is no $m_\gamma$-dependence on a spherically symmetric background.
\item We combined the $X,X'={\rm up,in}$ terms containing $\llbracket I^{++}_{{\rm up},k,{\rm in},k',{\rm up},\ell,(p)}(h,\omega - h,\omega) \rrbracket \llbracket I^{++ \ast}_{{\rm up},k,{\rm in},k',{\rm in},\ell,(p)}(h,\omega - h,\omega) \rrbracket$ with the in,up terms containing $\llbracket I^{++}_{{\rm in},k,{\rm up},k',{\rm up},\ell,(p)}(h,\omega - h,\omega) \rrbracket \llbracket I^{++ \ast}_{{\rm in},k,{\rm up},k',{\rm in},\ell,(p)}(h,\omega - h,\omega) \rrbracket$ using the symmetry relation (straightforwardly verified from Eq.~\ref{eq: coupling_int}): 
\begin{equation}
\llbracket I^{++}_{X,k,X',k',\ell, (p)}(h,h',\omega) \rrbracket =
(-1)^{j-j'} \llbracket I^{++}_{X',-k',X,-k,\ell, (p)}(h',h,\omega) \rrbracket.
\end{equation}
These terms are equivalent after we take the $\sum_{kk'}$ over both positive and negative values, and perform the fermion energy integral over the allowed range, $\int_0^\omega {dh}/({2\pi})$.
\item The two terms that have the same set of indices are in the interference term, $\Re T^\ast_{1,\ell,\omega} R_{1,\ell,\omega}$, correspond to the coupling $\llbracket I^{-+}_{{\rm in},k,{\rm up},k',{\rm up},\ell,(p)}(h,\omega + h,\omega) \rrbracket \llbracket I^{-+*}_{{\rm in},k,{\rm up},k',{\rm in},\ell,(p)}(h,\omega + h,\omega) \rrbracket $. They have been combined. Physically, this is because the interaction being described is of the form $e_{{\rm up}} \rightarrow \gamma_{\rm up/in} + e_{\rm in}$, where the $e$ can be either electron or position by $\mathcal{C}$ -- symmetry. However, the transition with $\gamma_{\rm in}$ corresponds to a process where the photon is \textit{reflected} back to the horizon and the process with $\gamma_{\rm up}$ correspond to the \textit{transmission} of a photon out to infinity. Since these two processes are identical in the electron/positron states, the photons will interfere, as expected by the combining of these two terms.
\end{list}

Each of the 13 coupling integral terms in Eq.~(\ref{eq: emitted_correction}) corresponds to a physical process involving a single-vertex interaction of $e^-$, $e^+$, and $\gamma$, with each particle in either the ``up'' or ``in'' mode. The channel the photon is in can be understood from the prefactor of either $T_{1,\ell,\omega}$ or $R_{1,\ell,\omega}$. For interactions with $|R_{1,\ell,\omega}|^2$, the 3 terms describe the processes:
\begin{equation}
e_{\rm up} \rightarrow e_{\rm in} + \gamma_{\rm in}, ~~~~ e_{\rm up} + e_{\rm up} \rightarrow \gamma_{\rm in}, ~~~~{\rm and}~~~~e_{\rm up} \rightarrow e_{\rm up} + \gamma_{\rm in},
\label{eq:list-R}
\end{equation}
respectively.
In these interactions, the factor of $|R_{1,\ell,\omega}|^2$ indicates these are the amplitudes processes that result in radiation emitted in the ``in'' mode (hence the label $\gamma_{\rm in}$) and reflected from the effective potential barrier so that the photons propagate out. (We label ``$e$'' here to denote particles of either sign charge.) Only emission terms are allowed here since there are no initial photons in the ``in'' modes.

Similar considerations apply to the 4 terms with $|T_{1,\ell,\omega}|^2$. These describe processes involving photons in the ``up'' mode ($\gamma_{\rm up}$) interacting with $e^-e^+$. These photons can be transmitted through the barrier and propagate out. Both emission and absorption processes are allowed here since there are initially photons in the ``up'' mode: the terms correspond respectively to
\begin{equation}
\gamma_{\rm up} \rightarrow e_{\rm in} + e_{\rm in},~~~~
e_{\rm up} \rightarrow e_{\rm in} + \gamma_{\rm up},~~~~
\gamma_{\rm up} \rightarrow e_{\rm up} + e_{\rm in},~~~~{\rm and}~~~~
e_{\rm up} + \gamma_{\rm up} \rightarrow e_{\rm in}.
\label{eq:list-T}
\end{equation}
The process $e_{\rm up} + e_{\rm up} \leftrightarrow \gamma_{\rm up}$ is allowed, but the net rate is zero since it is in thermodynamic equilibrium and so it does not appear in Eq.~(\ref{eq: emitted_correction}).

Finally, we have the terms with $\Re T^\ast_{1,\ell,\omega} R_{1,\ell,\omega}$, corresponding to the \textit{interference} of the ``in,reflected'' and ``out,transmitted'' processes since the initial and final state for the fermions are the same but with different photon modes. This is analogous to the familiar interference pattern created by a dipole antenna over a ground plane, in which the {\em amplitude} radiation pattern includes both the radiation emitted directly upward by the antenna and the the radiation emitted into the ground plane and reflected back up. The power radiation pattern then contains an interference pattern. Each of the 7 terms in Eqs.~(\ref{eq:list-R}) and (\ref{eq:list-T}) has a corresponding interference term, but there are only 6 terms with $\Re T^\ast_{1,\ell,\omega} R_{1,\ell,\omega}$ because the process $e_{up} \rightarrow e_{\rm in}+\gamma_{\rm in\,or\,up}$ appears in both Eqs.~(\ref{eq:list-R}) and (\ref{eq:list-T}), hence the interference terms can be combined.

Equation~(\ref{eq: emitted_correction}) is our final expression for the dissipative ${\cal O}(\alpha)$ correction to the Hawking radiation on a Schwarzschild metric. 

\section{Discussion}
\label{sec: Discussion}

Primordial black holes are novel objects that are both a dark matter candidate and a potential probe of epochs in the early Universe such as the late stages of inflation or phase transitions, which are difficult to probe with the ``standard'' measurements in cosmology. One point of observational interest is the Hawking radiation emitted from PBHs, since the lower end of the allowed mass range should emit in the $\gamma$-ray regime. This -- combined with reports that the formally ${\cal O}(\alpha)$ internal bremsstrahlung effect can qualitatively change the spectrum of Hawking radiation and impact detectability \cite{2021PhRvL.126q1101C} -- motivates accurate calculation of the emitted radiation in order to place constraints on PBHs. We are particularly interested in whether at order ${\cal O}(\alpha)$ there are any other qualitatively new features in the Hawking radiation.

This paper is the first in a series that aims to compute the full suite of corrections to the emitted radiation spectrum first order in the fine structure constant, $\alpha$, for a Schwarzschild black hole in a mass range where electrons become important ($T_{\rm H}/m_e$ of order unity). This requires performing a perturbative QED calculation with photon, electron, and positron interactions on a curved background. We follow the canonical quantization procedure starting from the QED action (Eq.~\ref{eq:action-QED}), then describe the vector potential and electron fields as sums over the creation and annihilation operators in Eqs.~(\ref{eq: vector_potential_final}) and (\ref{eq: psi_electron}). Then we compute the interaction Hamiltonian, Eq.~(\ref{eq: Hamiltonian_interact}), and rewrite it such that it can be easily translated to time-dependent perturbative techniques for computing evolution of operators (Eq.~\ref{eq: operator_evolution}) using the initial conditions with no radiation coming ``in'' from $\infty$ (black hole placed in vacuum) and thermal Hawking radiation coming up from the horizon (Eq.~\ref{eq: phase_densities}). We also tested our expressions for the interaction Hamiltonian, specifically Eq.~(\ref{eq: coupling_int}), with limiting cases of bound-state ($h<\mu$) electric and magnetic dipole transitions in the non-relativistic limit in flat spacetime ($|h-\mu|\ll\mu$ and $r\gg M$) in Appendix~\ref{app: test_case}, and found both cases to agree with standard relations from atomic physics. The evolution of the photon field is captured to ${\cal O}(\alpha)$ in Eq.~(\ref{eq: nested_1}), but we can see that there are two different mechanisms that can affect the Hawking radiation up to this order:
\begin{list}{$\bullet$}{}
\item The emission/absorption of photons from interactions with the spinor fields. These are {\em dissipative} effects, associated with the real and even function $\delta(\Omega)$ in Eq.~(\ref{eq: Phi_Omega}). Here $\Omega$ is the net energy change in the reactions of Eqs.~(\ref{eq:list-R}) and (\ref{eq:list-T}), and the $\delta$-function enforces conservation of energy.
\item The processes that do not change the total number of photons but rather change the barrier penetration probability, i.e., vacuum polarization and plasma frequency effects. These are {\em conservative} effects, associated with the imaginary and odd function $(i/\pi){\rm P}(1/\Omega)$ in Eq.~(\ref{eq: Phi_Omega}). Here $\Omega$ is the amount of energy that must be ``borrowed'' to go to a virtual state.
\end{list}
This paper is concerned with the analytic form of the dissipative sector of the evolution. We arrive ultimately at the final expression for the dissipative correction to the emitted radiation, Eq.~(\ref{eq: emitted_correction}). This is expressed as a mode sum (or integral, for the one continuous variable), with each term containing the square norm or complex product of the $I$-integrals, which describe the overlap of the Schwarzschild wave functions of the fermions and photons.

This paper focuses on the analytic formulation of Eq.~(\ref{eq: emitted_correction}), after using rotation, time translation, and parity symmetries to completely simplify the result. A forthcoming paper (Paper II) will focus on the numerical implementation of Eq.~(\ref{eq: emitted_correction}). The implementation is numerically challenging -- in addition to the large number of nested sums and integrals, the $I$-integrals converge slowly at large $r_\star$ (the integrands scale as an oscillatory function with logarithmically divergent phase corrections times $1/r$). Furthermore, at large $M$, massive fermions such as $e^\pm$ can be in classically stable orbits around the black hole. In wave mechanics, the classical turning points become a cavity, and there are resonances at energies where a half-integer number of fermion waves (in the WKB sense) fit across the cavity. The resonances must be properly sampled in the integration over energies $h$ and $h'$ (see, e.g., \citet{1972JPhB....5..277B} for a discussion of similar problems in radiative formation of molecules). We plan to present a treatment of these issues in Paper II.

This concludes the analytic part of the dissipative $\mathcal{O}(\alpha)$ correction to the Hawking radiation, but begins the start of several other avenues of work that will encapsulate different perturbative effects to improve the distribution of Hawking radiation from PBHs. The next $\mathcal{O}(\alpha)$ correction that was mentioned, but not taken into account in this analysis, is the conservative effect that affects the transmission coefficient. This includes both plasma effects and the vacuum polarization effect in the Schwarzschild spacetime. In flat spacetime, the ${\cal O}(\alpha)$ or 1-loop correction to the photon propagator due to vacuum polarization is divergent, requiring renormalization to produce finite results \cite{1995iqft.book.....P, 2000qtf..book.....W, 2007qft..book.....S}. The same thing should happen here: the conservative correction includes an integral with the principal part of $1/\Omega$, and the energy integral and sum over angular momentum modes together will give an infinite contribution. We intend to explore renormalization approaches in future work.

Other possible future works could extend these techniques to more complicated systems of interacting particles on black hole spacetimes. One such example is the production of pions at $T_{\rm H} \gtrsim 20$ MeV; these have strong self-interactions, and are a favorable mode of producing radiation since they are spin 0 (which leads to an $\ell=0$ channel with no angular momentum barrier) and their decay can lead to secondary gamma rays and positrons \cite{1990PhRvD..41.3052M, 1991PhRvD..44..376M}. Additionally, including the spin of the PBH -- that is, going from a Schwarzschild spacetime to a Kerr spacetime -- could be an interesting avenue for future work as this enhances production of higher spin particles \cite{2022MNRAS.517L...1T, 2022PhRvD.105a5022C}. The Kerr background introduces several novel effects due to the lack of spherical symmetry (so we cannot simplify the problem using the Wigner-Eckart theorem) and the lack of time-reversal symmetry (which results in the ``out+down'' vacuum being different from the ``in+up'' vacuum \cite{2013PhRvD..87f4027C}). These effects are all necessary for a complete and accurate description of radiative processes from PBHs, which is imperative for future interpretations of PBH-related implications on cosmology. 

\begin{acknowledgments}

Mahalo nui loa to Yuri Kovchegov, Heyang Long, Samir Mathur, Rachel Slaybaugh, and Todd Thompson for insightful and helpful discussions.

During the preparation of this work, the authors were supported by David \& Lucile Packard Foundation award 2021-72096, the Simons Foundation award 60052667, and NASA award 15-WFIRST15-0008.

\end{acknowledgments}

\appendix

\section{Some useful identities for the angular integrals}
\label{app:identities}

Here we tabulate some identities for the angular integrals $\Delta^{kk'l}_{mm'm_\gamma}$ and $\Pi^{kk'l\,\pm}_{mm'm_\gamma}$ defined in Eqs.~(\ref{eq: DELTA}) and (\ref{eq: PI}).

Our first identity follows from the recursion relation for Clebsch-Gordan coefficients, Eq.~(2.2.4) of \cite{1960amqm.book.....E}. If one writes this recursion relation for $m_1 = m_2 = 1/2$ and $m=0$, and expresses the result in terms of $3j$ symbols, then one obtains
\begin{equation}
\sqrt{j_3(j_3+1)} \tj{j_1}{j_2}{j_3}{\frac12}{\frac12}{-1} = -\left(j_1+\frac12\right) \tj{j_1}{j_2}{j_3}{-\frac12}{\frac12}{0}
-\left(j_2+\frac12\right) \tj{j_1}{j_2}{j_3}{\frac12}{-\frac12}{0}.
\end{equation}
Using the symmetry relation that one may interchange any two columns of a $3j$ symbol, with a factor of $(-1)^{j_1+j_2+j_3}$, we may combine the $3j$ symbols on the right-hand side. We may also use the symmetry relation that all of the $m$'s can flip sign with a factor of $(-1)^{j_1+j_2+j_3}$. This results in
\begin{equation}
\tj{j_1}{j_2}{j_3}{-\frac12}{-\frac12}{1}
= -\frac{1}{\sqrt{j_3(j_3+1)}}W_{j_1j_2j_3} \tj{j_1}{j_2}{j_3}{\frac12}{-\frac12}{0},
~~~
W_{j_1j_2j_3} = j_1+\frac12 + (-1)^{j_1+j_2+j_3}\left( j_2 + \frac12 \right)
.\label{eq:TJW}
\end{equation}
Comparing this to Eqs.~(\ref{eq: DELTA}) and (\ref{eq: PI}), we find that $\Pi^{kk'\ell\,-}_{mm'm_\gamma}$ and $\Delta^{kk'\ell}_{mm'm_\gamma}$ are only non-zero if $(-1)^{j+j'+\ell}=-ss'$. We therefore have $W_{jj'\ell} = |k| + ss'|k'| = s(k-k')$. Noting that the difference between $\Pi^{kk'\ell\,-}_{mm'm_\gamma}$ and $\Delta^{kk'\ell}_{mm'm_\gamma}$ is this $3j$ symbol, a prefactor of $-s$, and the factor of 2, we find that
\begin{equation}
\Pi^{kk'\ell\,-}_{mm'm_\gamma} = 2\frac{k-k'}{\sqrt{\ell(\ell+1)}}  \Delta^{kk'\ell}_{mm'm_\gamma}.
\label{eq:Pi-Delta}
\end{equation}
Thus the two angular integrals for the $(e)$ harmonics -- the $\Delta^{kk'\ell}_{mm'm_\gamma}$ integrals that come from the radial parts of the vector potential and the $\Pi^{kk'\ell\,-}_{mm'm_\gamma}$ integrals that come from the angular parts -- in fact are closely related.

A corollary to Eq.~(\ref{eq:TJW}) is that $\Pi^{kk'\ell+}_{mm'm_\gamma}$ can be expressed in terms of the $3j$ symbols with $m$-values $\frac12,-\frac12,0$. We note that $\Pi^{kk'\ell+}_{mm'm_\gamma}$ has a selection rule that $(-1)^{j+j'+\ell}=ss'$, so that $W_{jj'\ell} = |k| + ss'|k'| = s(k+k')$. Therefore,
\begin{equation}
\Pi^{kk'\ell+}_{mm'm_\gamma} = (-1)^{m+1/2} \sqrt{\frac{(2j+1)(2j'+1)(2\ell+1)}{4\pi\ell(\ell+1)}}
(k+k')
\tj{j}{j'}{\ell}{-m}{m'}{m_\gamma}
\tj{j}{j'}{\ell}{\frac12}{-\frac12}{0}
(1 - ss'(-1)^{j-j'+\ell}).
\label{eq:Pi-Plus-Reduce}
\end{equation}

\section{Expectation values of four fermion operators}
\label{app: four_fermion_op}

We need the expectation values of different combinations of four fermion operators that appear in Eq.~(\ref{eq: operator_evolution}) from products of the form $\langle Q^{(p)}_{X_{\gamma}\beta}Q^{ (p)\dagger}_{X'_{\gamma}\beta} \rangle$. These are needed only in the unperturbed case where all of the modes are independent. These are:
\begin{eqnarray}
\label{eq: four_fermion_operators}
\langle \hat b^\dagger_{Xkmh} \hat d^\dagger_{X'k'm'h'} \hat d_{X''k''m''h''}  \hat b_{X'''k'''m'''h'''} \rangle &=& 2\pi f^{e^-}_{XX'''km}(h''') \delta(h-h''') \delta_{kk'''} \delta_{mm'''} 
\nonumber \\ && ~~~~\times
2\pi f^{e^+}_{X'X''k'm'}(h'') \delta(h'-h'') \delta_{k'k''} \delta_{m'm''},
\nonumber \\
\langle \hat d_{Xkmh} \hat b_{X'k'm'h'} \hat b^\dagger_{X''k''m''h''} \hat d^\dagger_{X'''k'''m'''h'''} \rangle &=& 2\pi g^{e+}_{XX'''km}(h''') \delta(h-h''') \delta_{kk'''} \delta_{mm'''} 
\nonumber \\ && ~~~~\times
2\pi g^{e^+}_{X'X''k'm'}(h'') \delta(h'-h'') \delta_{k'k''} \delta_{m'm''},
\nonumber \\
\langle \hat b^\dagger_{Xkmh} \hat b_{X'k'm'h'} \hat d_{X''k''m''h''} \hat d^\dagger_{X'''k'''m'''h'''} \rangle &=& 2\pi f^{e^-}_{XX'kmh}(h')\delta(h-h')\delta_{kk'} \delta_{mm'}
\nonumber \\ && ~~~~\times
2\pi g^{e+}_{X''X'''k''m''}(h''')\delta(h''-h''') \delta_{k''k'''} \delta_{m''m'''},
\nonumber \\
\langle \hat d_{Xkmh} \hat d^\dagger_{X'k'm'h'} \hat b^\dagger_{X''k''m''h''} \hat b_{X'''k'''m'''h'''} \rangle &=& 2\pi g^{e^+}_{XX'km}(h')\delta(h-h') \delta_{kk'} \delta_{mm'}
\nonumber \\ && ~~~~\times
2\pi f^{e^-}_{X''X'''k''m''}(h''') \delta(h'' -h''') \delta_{k''k'''} \delta_{m''m'''},
\nonumber \\
\langle \hat b^\dagger_{Xkmh} \hat b_{X'k'm'h'} \hat b^\dagger_{X''k''m''h''} \hat b_{X'''k'''m'''h'''} \rangle &=& 2\pi f^{e^-}_{XX'km}(h') \delta(h-h') \delta_{kk'} \delta_{mm'} 
\nonumber \\ && ~~~~\times
2\pi f^{e^-}_{X''X'''k''m''}(h''') \delta(h''-h''') \delta_{k''k'''} \delta_{m''m'''}  
\nonumber \\ && +
2\pi f^{e^-}_{XX'''km}(h''') \delta(h-h''') \delta_{kk'''} \delta_{mm'''}
\nonumber \\ && ~~~~\times
2\pi g^{e^-}_{X'X''km}(h'') \delta(h'-h'') \delta_{k'k''} \delta_{m'm''}, ~~~{\rm and}
\nonumber \\
\langle \hat d_{Xkmh} \hat d^\dagger_{X'k'm'h'} \hat d_{X''k''m''h''} \hat d^\dagger_{X'''k'''m'''h'''} \rangle &=& 2\pi g^{e^+}_{XX'km}(h') \delta(h-h') \delta_{kk'} \delta_{mm'}
\nonumber \\ && ~~~~\times
2\pi g^{e^+}_{X''X'''k''m''}(h''') \delta(h''-h''') \delta_{k''k'''} \delta_{m''m'''} 
\nonumber \\ && +
2\pi g^{e^+}_{XX'''km}(h''') \delta(h-h''') \delta_{kk'''} \delta_{mm'''}  
\nonumber \\ && ~~~~\times
2\pi f^{e^+}_{X'X''k'm'}(h'') \delta(h'-h'') \delta_{k'k''} \delta_{m'm''}.
\end{eqnarray}
We can derive the first four expressions in Eq.~(\ref{eq: four_fermion_operators}) by assuming that the electron and positron fields are uncorrelated, i.e. one can group and separate expectation values of combinations of $\hat b$ and $\hat d$ (note that this grouping produces an even number of $-$ signs from fermion operator anticommutation). For the last two expressions -- the expectation values with four electron or four positron operators -- we can directly verify the results if the density matrix is diagonal in the Fock space. Since the two sides of these equations transform the same way under unitary transformations on the $X,X',...$ indices, they remain valid in another choice of basis.

\section{Relating the ``In/Up'' scattering basis to the ``Out/Down'' scattering basis for the photon}
\label{app:out-down}

We wish to relate the current basis set of solutions, Eqs.~(\ref{eq:psi-in}) and (\ref{eq:psi-up}), of the photon radial equation to the set of solutions that describe the radiation going ``down/out'' of the black hole. We will also assume that the black hole is in a vacuum and there is no infalling radiation from an external source.

Since Eq.~(\ref{eq:HL}) is a second order ordinary differential equation, it will have two independent solutions. Therefore, we can define any other set of solutions as linear combinations of the original two independent solutions. This allows us to define the asymptotic ``out/down'' solutions as
\begin{equation}
\label{eq: out_down1}
\begin{pmatrix}
\Psi_{\ell,\omega, \rm out} \\
\Psi_{\ell,\omega, \rm down}
\end{pmatrix} =
\begin{pmatrix}
A & B \\
C & D
\end{pmatrix}
\begin{pmatrix}
\Psi_{\ell,\omega, \rm up} \\
\Psi_{\ell,\omega, \rm in}
\end{pmatrix}
\end{equation}
where
\begin{equation}
\Psi_{\ell,\omega,\rm (out,down)}(r_\star) \rightarrow
\left\{ \begin{array}{lcl} (A,C)e^{i\omega r_\ast} + ((B,D)T_{1,\ell, \omega} - (A,C)R^\ast_{1,\ell,\omega}e^{2i{\rm arg}T_{1,\ell,\omega}})e^{-i\omega r_\ast} & ~ & r_\star\rightarrow -\infty \\
((A,C)T_{1,\ell,\omega} + (B,D)R_{1,\ell,\omega})e^{i\omega r_\ast} + (B,D)e^{-i\omega r_\ast} & & r_\star\rightarrow\infty
\end{array}\right.
\label{eq:out_down2}
\end{equation}
where $A$, $B$, $C$, and $D$ are coefficients relating the two different sets of solutions.

In order to preserve the same normalizations used in the ``up/in'' solutions, we can constrain the coefficients that appear in the linear transformation Eq.~(\ref{eq: out_down1}). For the outgoing solution ($\Psi_{\rm out}$), we want the boundary condition similar to the $\Psi_{\rm up}$: a purely outgoing wave coming from the black hole and a superposition of outgoing and ingoing waves far from the horizon. Similarly for the down going solution ($\Psi_{\rm down}$): a purely ingoing wave far from the black hole and a superpostion of ingoing and outgoing waves near the horizon. Thus, the linear transformation that satisfies these boundary conditions (normalization conventions) are
\begin{equation}
\label{eq: change_basis_photon}
\begin{pmatrix}
\Psi_{\ell,\omega, \rm out} \\
\Psi_{\ell,\omega, \rm down}
\end{pmatrix} =
\begin{pmatrix}
T^\ast_{1,\ell, \omega} & R^\ast_{1,\ell,\omega} \\
-R_{1,\ell,\omega}e^{-2i{\rm arg}T_{1,\ell,\omega}} & T^\ast_{1,\ell,\omega}
\end{pmatrix}
\begin{pmatrix}
\Psi_{\ell,\omega, \rm up} \\
\Psi_{\ell,\omega, \rm in}
\end{pmatrix},
\end{equation}
where the transformation matrix is unitary, which we expected due to conservation of probability, and has determinant $e^{-2i{\rm arg}T_{1,\ell,\omega}}$. 

Now with this transformation, we can compute the intensity of radiation leaving the black hole i.e. $f^\gamma_{{\rm out,out},\ell m_{\gamma}(p)}(\omega)$. Since the vector potentials, Eq.~(\ref{eq: vector_potential_final}), are sums over the solutions to the radial photon equation (sums over ``in/up'' states), we can expand the $\Psi_{X_{\gamma} \ell \omega}$ into ``out/down'' states using Eq.~(\ref{eq: change_basis_photon}) and then collecting the operators corresponding the $\Psi_{{\rm out/down}}$. This leads to the expressions
\begin{eqnarray}
\hat a_{{\rm out}, \ell m_{\gamma} \omega(p)} &=& R_{1,\ell,\omega} \hat a_{{\rm in}, \ell m_{\gamma} \omega (p)} + T_{1,\ell,\omega} \hat a_{{\rm up}, \ell m_{\gamma} \omega (p)}
~~~{\rm and}
 \nonumber \\ 
\hat a_{{\rm down}, \ell m_{\gamma} \omega(p)} &=& T_{1,\ell,\omega} \hat a_{{\rm in}, \ell m_{\gamma} \omega (p)} - e^{2i{\rm arg}T_{1,\ell,\omega}}R^\ast_{1,\ell,\omega} \hat a_{{\rm up}, \ell m_{\gamma} \omega (p)}.
\end{eqnarray}
Now the outgoing phase space density is
\begin{eqnarray}
2\pi\delta(\omega-\omega')\,
f^\gamma_{{\rm out,out},\ell m_{\gamma}(p)}(\omega) &=& \langle \hat a^\dagger_{{\rm out},\ell m_{\gamma} \omega (p)} \hat a_{{\rm out},\ell m_{\gamma} \omega' (p)}\rangle
\nonumber \\
&=& \langle (R^\ast_{1,\ell,\omega} \hat a^\dagger_{{\rm in}, \ell m_{\gamma} \omega (p)} + T^\ast_{1,\ell,\omega} \hat a^\dagger_{{\rm up}, \ell m_{\gamma} \omega (p)})(R_{1,\ell,\omega} \hat a_{{\rm in}, \ell m_{\gamma} \omega' (p)} + T_{1,\ell,\omega} \hat a_{{\rm up}, \ell m_{\gamma} \omega' (p)}) \rangle 
\nonumber \\ &=&
|R_{1, \ell, \omega}|^2 \langle \hat a^\dagger_{{\rm in}, \ell m_{\gamma} \omega (p)} a_{{\rm in}, \ell m_{\gamma} \omega' (p) } \rangle +  
|T_{1,\ell, \omega}|^2\langle \hat a^\dagger_{{\rm up}, \ell m_{\gamma} \omega (p)} \hat a_{{\rm up}, \ell m_{\gamma} \omega' (p)} \rangle 
\nonumber \\ &&
+ T_{1,\ell,\omega} R^\ast_{1, \ell, \omega} \langle \hat a^\dagger_{{\rm in}, \ell m_{\gamma} \omega (p)} \hat a_{{\rm up}, \ell m_{\gamma} \omega' (p) } \rangle
+ T^\ast_{1,\ell,\omega} R_{1, \ell, \omega} \langle \hat a^\dagger_{{\rm up}, \ell m_{\gamma} \omega (p)} \hat a_{{\rm in}, \ell m_{\gamma} \omega' (p) } \rangle.
\label{eq: out_out_intensity}
\end{eqnarray}

\section{Test case: electric dipole emission from non-relativistic bound electrons}
\label{app: test_case}

This appendix considers the emission of photons from non-relativistic bound electrons emitted from the black hole as a test case to check the normalization factors in Eq.~(\ref{eq: evolve_diss}). This is analogous to the case of dipole radiation from a hydrogenic atom, since both the gravitational and electrostatic interactions have a $1/r$ potential (so $M\mu$ replaces the $Z/137$ of atomic physics). In the case of the black hole, the occupation probability of each ``orbital'' is given by the Hawking blackbody factor $1/(e^{8\pi Mh}+1)$ (although this turns out not to matter for this test). The approximations involved are:
\begin{itemize}
\item Non-relativistic: $|h-\mu| \ll \mu$. (We focus on the bound case, $h<\mu$.)
\item Emission far enough from the black hole ($r\gg M$).
\item Low-frequency dipole radiation, $\omega \ll 1/r$.
\end{itemize}
We consider both electric and magnetic dipole radiation.

\subsection{Electric dipole transitions: $\ell = 1~\&~(p) = (e)$}
Since we work far from the black hole, we neglect the difference between $\partial_{r_\star}$ and $\partial_r$. We focus on the leading (electric dipole) emission, i.e., $\ell=1$ and even ($p=e$) parity. The bound case ($h<\mu$) implies that there are only ``up'' modes for the electron, no ``in'' modes.

The electron wave equation in this case simplifies to
\begin{equation}
hF = \mu\left(1-\frac Mr\right)F + \frac krG + \partial_rG
~~~{\rm and}~~~
hG = \frac krF - \partial_rF - \mu\left(1-\frac Mr\right)G.
\end{equation} 
The latter equation simplifies to
\begin{equation}
G = \frac1{2\mu} \left(\frac kr - \partial_r \right)F,
\label{eq:app-G}
\end{equation}
which, plugged into the first equation, gives
\begin{equation}
\frac12 \mu v^2 F = -\frac{\mu M}r F + \frac{k(k-1)}{2\mu r^2}F - \frac1{2\mu}\partial_r^2F.
\label{eq:app-F}
\end{equation}
Following standard atomic physics notation, we describe $F$ as the ``large component'' and $G$ as the ``small component.'' We see that $F$ satisfies the usual Schr\"odinger equation, with the usual orbital angular momentum $L = j-\frac12s$ so that $L(L+1)=k(k-1)$.

The photon wave function in this limit is the solution with only the angular momentum barrier,
\begin{equation}
\label{eq: app-photon}
\Psi_{{\rm in},1,\omega}(r) = -2 \omega r j_1(\omega r) \rightarrow -\frac23\omega^2r^2 + {\cal O}(r^4),
\end{equation}
where $j_1$ is the spherical Bessel function. The barrier transmission coefficient $T_{1,1,\omega}$ can be neglected since $\omega M\ll 1$.

For the $3j$ symbol that will appear in the overlap integral, we have
\begin{equation}
\sum_{mm'} |\Delta^{kk'1}_{mm'm_\gamma}|^2 = \frac{(2j+1)(2j'+1)}{4\pi} \left(\begin{array}{ccc} j & j' & 1 \\ \frac12 & -\frac12 & 0 \end{array}\right)^2 \delta_{-ss',(-1)^{j-j'}}.
\end{equation}

We then compute the integral in $I^{+-}_{{\rm up},k,m,{\rm up},k',m',{\rm in},1,m_\gamma,(e)}(h,h',\omega)$ (in this equation, the ``in'' and ``up'' subscripts are implied, and we have used two integrations by parts, and used the boundary condition that the wave function goes to zero at $r=\infty$):
\begin{eqnarray}
&& \!\!\!\!\!\!\!\!\!\!\!\!\!\!\!\!
\int_0^\infty \Bigl[(F^\ast_{kh}G_{k'h'} - G^\ast_{kh}F_{k'h'})\Psi_{1\omega} \frac1{r^2\omega^{3/2}}
+ (F^\ast_{kh}G_{k'h'} + G^\ast_{kh}F_{k'h'})\Psi'_{1\omega} \frac{k-k'}{2r\omega^{3/2}}
\Bigr] dr
\nonumber \\
&=& - \frac23\omega^{1/2} \int_0^\infty \Bigl[ 
F^\ast_{kh}G_{k'h'} - G^\ast_{kh}F_{k'h'}
+ (k-k')(F^\ast_{kh}G_{k'h'} + G^\ast_{kh}F_{k'h'})
\Bigr] dr
\nonumber \\
&=& - \frac{\omega^{1/2}}{3\mu} \int_0^\infty \Bigl[ 
\frac{k'-k}{r} F^\ast_{kh}F_{k'h'}
+ \frac{k^2-k'{^2}}{r} F^\ast_{kh}F_{k'h'}
+(k'-k) \partial_r(F^\ast_{kh}F_{k'h'})
- [F^\ast_{kh}\partial_rF_{k'h'} - (\partial_r F^\ast_{kh})F_{k'h'}]
\Bigr] dr
\nonumber \\
&=& - \frac{\omega^{1/2}}{3\mu} \int_0^\infty \Bigl[ 
\frac{(k'-k)(1-k'-k)}{r} F^\ast_{kh}F_{k'h'}
+ r\partial_r [F^\ast_{kh}\partial_rF_{k'h'} - (\partial_r F^\ast_{kh})F_{k'h'}]
\Bigr] dr
\nonumber \\
&=& - \frac{\omega^{1/2}}{3\mu} \int_0^\infty \Bigl[ 
\frac{(k'-k)(1-k'-k)}{r} F^\ast_{kh}F_{k'h'}
+ r [F^\ast_{kh}\partial_r^2F_{k'h'} - (\partial_r^2 F^\ast_{kh})F_{k'h'}]
\Bigr] dr
\nonumber \\
&=& - \frac{\omega^{1/2}}{3\mu} \int_0^\infty \Bigl[ 
\frac{(k'-k)(1-k'-k)}{r} F^\ast_{kh}F_{k'h'}
+ r\left( 2\mu(h-h') + \frac{k'(k'-1)-k(k-1)}{r^2} \right) F^\ast_{kh}F_{k'h'}
\Bigr] dr
\nonumber \\
&=& -\frac23 (h-h')  \omega^{1/2} \int_0^\infty r  F^\ast_{kh}F_{k'h'}\, dr.
\end{eqnarray}

Now the overlap integral corresponding to an electron dropping to a lower-energy state is:
\begin{eqnarray}
\sum_{mm'} |I^{+-}_{{\rm up},k,m,{\rm up},k',m',{\rm in},1,m_\gamma,(e)}(h,h',\omega)|^2 &=&
 \frac{(h-h')^2\omega}{9\mu^2}
 \frac{(2j+1)(2j'+1)}{4\pi} \left(\begin{array}{ccc} j & j' & 1 \\ \frac12 & -\frac12 & 0 \end{array}\right)^2 \delta_{-ss',(-1)^{j-j'}}
\nonumber \\ &&
\times
  \left| \int_0^\infty r  F^\ast_{kh}F_{k'h'}\, dr \right|^2.
\end{eqnarray}
The $3j$ symbols can be evaluated in each case (see the tables in Edmonds \cite{1960amqm.book.....E}):
\begin{eqnarray}
\sum_{mm'} |I^{+-}_{{\rm up},k,m,{\rm up},k',m',{\rm in},1,m_\gamma,(e)}(h,h',\omega)|^2 &=&
 \frac{(h-h')^2\omega}{144\pi\mu^2} \times
\left\{
\begin{array}{lcl}
\frac{(2j+1)(2j+3)}{j+1} & & j'=j+1 \\
\frac{2j+1}{j(j+1)} & & j'=j \\
\frac{(2j-1)(2j+1)}{j} & & j'=j-1 \\
0 & & {\rm otherwise}
\end{array}
\right\}
\nonumber \\ && 
\times \delta_{-ss',(-1)^{j-j'}}
  \left| \int_0^\infty r  F^\ast_{kh}F_{k'h'}\, dr \right|^2.
\end{eqnarray}

If we sum over the final value of $s'$ at fixed $L'$, and average over initial states (so dividing by $2j+1$), we get:
\begin{eqnarray}
\frac1{2j+1}
\sum_{mm's'} |I^{+-}_{{\rm up},k,m,{\rm up},k',m',{\rm in},1,m_\gamma,(e)}(h,h',\omega)|^2 &=&
 \frac{(h-h')^2\omega}{36\pi\mu^2} \times
\left\{
\begin{array}{lcl}
\frac{L+1}{2L+1} & & L'=L+1 \\
\frac{L}{2L+1} & & L'=L-1 \\
0 & & {\rm otherwise}
\end{array}
\right\}
\times
  \left| \int_0^\infty r  F^\ast_{kh}F_{k'h'}\, dr \right|^2.
  \nonumber \\ &&
\end{eqnarray}
The total emitted photon rate if $T_{1,1,\omega}\approx 0$ and $|R_{1,1,\omega}|\approx 1$ is then $2\ell+1=3$ times Eq.~(\ref{eq:ddt-out}),
including the relevant contribution from Eq.~(\ref{eq: nested_1}):
\begin{eqnarray}
\frac{d N_\gamma^{(1)}}{d\omega\,dt} &\ni& \frac 3{2\pi} e^2 \int \frac{dh\,dh'}{(2\pi)^2}
\sum_{kmk'} f^{e^-}_{{\rm up},{\rm up},k,h}g^{e^-}_{{\rm up},{\rm up},k',h'}
 \frac{(h-h')^2\omega}{36\pi\mu^2} \times
\left\{
\begin{array}{lcl}
\frac{L+1}{2L+1} & & L'=L+1 \\
\frac{L}{2L+1} & & L'=L-1 \\
0 & & {\rm otherwise}
\end{array}
\right\} 
\nonumber \\ && \times 
 \left| \int_0^\infty r  F^\ast_{kh}F_{k'h'}\, dr \right|^2
2\pi\delta(\omega+h'-h).
\label{eq:tot-1}
\end{eqnarray}
Here ``$\ni$'' indicates that the emitted photon rate includes this term; the term includes only spontaneous emission from bound electron levels to bound electron levels. Stimulated emission, absorption, and any processes involving positrons or unbound ($h>\mu$) electrons are excluded from Eq.~(\ref{eq:tot-1}). The photon emission rate can be simplified by integrating over frequency to
\begin{equation}
\frac{d N_\gamma^{(1)}}{dt} \ni e^2 \int \frac{dh\,dh'}{(2\pi)^2}
\sum_{kmk'} f^{e^-}_{{\rm up},{\rm up},h}g^{e^-}_{{\rm up},{\rm up},h'}
 \frac{(h-h')^3}{12\pi\mu^2} \times
\left\{
\begin{array}{lcl}
\frac{L+1}{2L+1} & & L'=L+1 \\
\frac{L}{2L+1} & & L'=L-1 \\
0 & & {\rm otherwise}
\end{array}
\right\} 
\times \left| \int_0^\infty r  F^\ast_{kh}F_{k'h'}\, dr \right|^2.
\label{eq:tot-1b}
\end{equation}

Equation~(\ref{eq:tot-1b}) can be contrasted with the usual rule for the emission rate from the usual dipole formula in non-relativistic quantum mechanics (see, e.g., Eq.~10.28b of \citet{1986rpa..book.....R}, with a factor of $4\pi$ in the denominator for conversion from the Gaussian system to units where $\epsilon_0=1$):
\begin{equation}
\frac{d N_\gamma^{(1)}}{dt} = e^2 \sum_{LM_LM_ShL'M'_LM'_Sh'} \delta_{M_SM'_S}
f^{e^-}_{{\rm up},{\rm up},h}g^{e^-}_{{\rm up},{\rm up},h'}
\frac{(h-h')^3}{3\pi}\left| \int_0^\infty {\boldsymbol r} \psi^\ast_{LM_Lh}({\boldsymbol r}) \psi_{L'M'_Lh'}({\boldsymbol r}) \, d^3{\boldsymbol r} \right|^2,
\end{equation}
where the sum is over all upper and lower states, we have expressed this sum in the $|M_L,M_S\rangle$ basis instead of the $|jm\rangle$ basis (these are related by Clebsch-Gordan coefficients), the $f$ and $g$ represent the probabilities for the upper state to be occupied and the lower state to be unoccupied, $h-h'$ is the natural frequency of the emitted photon, and the last object is the usual electric dipole matrix element. If the wave function is written as $\psi_{LM_Lh}(r,\theta,\phi) = R_{Lh}(r)Y_{LM_L}(\theta,\phi)/r$, we may perform the usual simplification of the spherical harmonic integrals,
\begin{eqnarray}
 \sum_{M'_L=-L'}^{L'} \left| \int_{S^2} Y^\ast_{LM_L}(\hat{\boldsymbol n}) \hat{\boldsymbol n} Y_{L'M'_L}(\hat{\boldsymbol n}) \,d^2\hat{\boldsymbol n} \right|^2
&=&
\frac{4\pi}{3} \sum_{M''=-1}^1 \sum_{M'_L=-L'}^{L'} \left| \int_{S^2} Y^\ast_{LM_L}(\hat{\boldsymbol n}) Y_{1M''}(\hat{\boldsymbol n}) Y_{L'M'_L}(\hat{\boldsymbol n}) \,d^2\hat{\boldsymbol n} \right|^2
\nonumber \\
&=& \left\{
\begin{array}{lcl}
\frac{L+1}{2L+1} & & L'=L+1 \\
\frac{L}{2L+1} & & L'=L-1 \\
0 & & {\rm otherwise}
\end{array}
\right.
\end{eqnarray}
(this follows from the 3-spherical harmonic integral and the $3j$ symbols in the table in Edmonds \cite{1960amqm.book.....E}).
Then we find
\begin{equation}
\frac{d N_\gamma^{(1)}}{dt} = e^2 \sum_{LM_LM_ShL'M'_Lh'} 
f^{e^-}_{{\rm up},{\rm up},h}g^{e^-}_{{\rm up},{\rm up},h'}
\frac{(h-h')^3}{3\pi}\times
\left\{
\begin{array}{lcl}
\frac{L+1}{2L+1} & & L'=L+1 \\
\frac{L}{2L+1} & & L'=L-1 \\
0 & & {\rm otherwise}
\end{array}
\right\} \times
\left| \int_0^\infty r R^\ast_{Lh}({ r}) R_{L'h'}({ r}) \, d{ r} \right|^2.
\label{eq:tot-2}
\end{equation}
In order to compare this to Eq.~(\ref{eq:tot-1b}), we recall that the radial wave functions in Eq.~(\ref{eq:tot-2}) are normalized to $\int |R^\ast_{Lh}|^2\,dr = 1$ with discrete energy levels, whereas the $F$'s are normalized by Eq.~(\ref{eq:FFGG}) with continuous energy levels. We may make a correspondence between these by defining a small energy level spacing $\Delta\varepsilon$, so Dirac delta functions in energy $\delta(h-h') \rightarrow \delta_{hh'}/\Delta\varepsilon$. The conclusion is that to go from Eq.~(\ref{eq:tot-1b}) to Eq.~(\ref{eq:tot-2}), we should make the replacements:
\begin{equation}
\label{eq:dipole_approx}
\int \frac{dh\,dh'}{(2\pi)^2} \rightarrow \frac{\Delta\varepsilon^2}{(2\pi)^2} \sum_{hh'}
~~~{\rm and}~~~
F_{kh} \rightarrow e^{i\varsigma}\sqrt{\frac{4\pi\mu}{\Delta\varepsilon}}\, R_{Lh},
\end{equation}
where we have used the approximation $|h|\approx\mu$, and $e^{i\varsigma}$ is an irrelevant phase factor. We see that with these replacements, Eqs.~(\ref{eq:tot-1b}) and (\ref{eq:tot-2}) are equivalent, as expected.

\subsection{Magnetic dipole transitions: $\ell = 1~\&~ (p) = (o)$}

Similar to the electric dipole transition, we can compute the limiting case for magnetic dipole transitions. The electron and photon wave function used above (Eq.~(\ref{eq:app-F})~\&~Eq.(\ref{eq: app-photon})) apply to this calculation, with the coupling integral in Eq.~(\ref{eq: coupling_int}) uses the odd parity sector for the electron occupying a lower energy state ($I^{+-}_{{\rm up},km,{\rm up},k'm',{\rm in}, 1m_\gamma,(o)}(h,h',\omega)$).

First we will need the square of the angular piece of the odd sector
\begin{equation}
\label{eq: square_odd_angle}
\sum_{mm'}|\Pi^{kk'1+}_{mm'm_\gamma}|^2 = \frac{4(2j+1)(2j'+1)}{4\pi}\tj{j}{j'}{1}{-1/2}{-1/2}{1}^2 \delta_{ss',(-1)^{j-j'}}. 
\end{equation}

Then we can expand the coupling integral to give
\begin{eqnarray}
\label{eq: overlap_odd}
\frac{1}{2\mu}\int^\infty_0 dr (F^\ast_{kh}G_{k'h'} + G^\ast_{kh}F_{k'h'})\Psi_{1,\omega}(r)\frac{1}{2r\sqrt{2\omega}}
&=& \frac{-\omega^{3/2}}{12\mu^{2}\sqrt{2}}\int^\infty_0 dr r(F^\ast_{kh}(\frac{k'}{r} - \partial_r)F_{k'h'} + F_{k'h'}(\frac{k}{r} - \partial_r)F^\ast_{kh})
\nonumber \\
&=& \frac{-\omega^{3/2}}{12 \mu^{2} \sqrt{2}}\int^\infty_0 dr (k+k'+1)F^\ast_{kh}F_{k'h'}.
\end{eqnarray}

We can combine these results to compute
\begin{eqnarray}
\label{eq: magnetic_transI}
\sum_{mm'}|I^{+-}_{{\rm up},km,{\rm up},k'm',{\rm in}, 1m_\gamma,(o)}(h,h',\omega)|^2 = \frac{\omega^3}{72 \mu^4}(k+k'+1)^2 \left|\int^\infty_0 dr F^\ast_{kh}F_{k'h'} \right|^2 \times
\nonumber \\
\frac{(2j+1)(2j'+1)}{4\pi}\tj{j}{j'}{1}{-1/2}{-1/2}{1}^2 \delta_{ss',(-1)^{j-j'}}.
\end{eqnarray}

The 3j--symbol can be evaluated using the tables in \citet{1960amqm.book.....E} to give
\begin{eqnarray}
\label{eq: magnetic_transII}
\sum_{mm'}|I^{+-}_{{\rm up},km,{\rm up},k'm',{\rm in}, 1m_\gamma,(o)}(h,h',\omega)|^2 = \frac{\omega^3}{2304 \pi \mu^4}(k+k'+1)^2 \left|\int^\infty_0 dr F^\ast_{kh}F_{k'h'} \right|^2 \times
\nonumber \\
\left\{
\begin{array}{lcl}
\frac{(2j+1)(2j+3)}{j+1} & & j'=j+1 \\
\frac{(2j+1)^3}{j(j+1)} & & j'=j \\
\frac{(2j-1)(2j+1)}{j} & & j'=j-1 \\
0 & & {\rm otherwise}
\end{array}
\right\} \delta_{ss',(-1)^{j-j'}}.
\end{eqnarray}

Like in the electric dipole case, if we sum over the final value of $s^{'}$ at a fixed $L^{'}=L$ and note that the only two terms that contribute are when $j^{'}=j \pm 1$ and which corresponds with $s^{'}= s \pm 2$ and average over initial states, we get:
\begin{eqnarray}
\label{eq: magnetic_transIII}
\frac{1}{2j+1}\sum_{mm's^{'}}|I^{+-}_{{\rm up},km,{\rm up},k'm',{\rm in}, 1m_\gamma,(o)}(h,h',\omega)|^2 = \frac{\omega^3}{144 \pi \mu^4} \left\{
\begin{array}{lcl}
\frac{L}{2L+1}  & & s=1, s^{'}=-1  \\ 
\frac{L+1}{2L+1} & & s=-1, s^{'}=1 \\ 
0 & & {\rm otherwise}
\end{array} \right\}
 \left|\int^\infty_0 dr F^\ast_{kh}F_{k'h'} \right|^2 
\end{eqnarray}
If we take the total emitted photon rate with the conditions of $T_{1,1,\omega} \approx 0$ and $\vert R_{1,1,\omega} \vert \approx 1$ along with averaging over the orientation which will give a factor of $2\ell+1=3$, then as in the case of the electric dipole we find:
\begin{eqnarray}
\frac{d N_\gamma^{(1)}}{d\omega\,dt} &\ni&  \frac{3e^{2}}{2 \pi} \int \frac{dh\,dh'}{(2\pi)^2}
\sum_{kmk'} \frac{\omega^3}{144 \pi \mu^4} \times f^{e^-}_{{\rm up},{\rm up},h}g^{e^-}_{{\rm up},{\rm up},h'} \left\{\begin{array}{lcl}
\frac{L}{2L+1}  & & s=1, s^{'}=-1  \\ 
\frac{L+1}{2L+1} & & s=-1, s^{'}=1 \\ 
0 & & {\rm otherwise}
\end{array} \right\}
\nonumber \\ && \times 
 \left| \int_0^\infty  F^\ast_{kh}F_{k'h'}\, dr \right|^2
2\pi\delta(\omega+h'-h).
\label{eq:dipole-1}
\end{eqnarray}
This term can be classified as a contribution from spontaneous emission. Terms related to stimulated emission, absorption, or any other process involving positrons or unbound electrons can be excluded. After integrating over frequency, we arrive at
\begin{eqnarray}
\label{eq:mag_dipole}
\frac{d N_\gamma^{(1)}}{dt} &\ni& e^2 \int \frac{dh\,dh'}{(2\pi)^2}
\sum_{kmk'} \frac{(h-h')^3}{48 \mu^4 \pi} \times f^{e^-}_{{\rm up},{\rm up},h}g^{e^-}_{{\rm up},{\rm up},h'}
\left\{
\begin{array}{lcl}
\frac{L}{2L+1}  & & s=1, s^{'}=-1  \\ 
\frac{L+1}{2L+1} & & s=-1, s^{'}=1 \\ 
0 & & {\rm otherwise}
\end{array} \right\}
 \left| \int_0^\infty  F^\ast_{kh}F_{k'h'}\, dr \right|^2.
\end{eqnarray}

As in the case of the electric dipole, to gain a physical intuition towards Eq.~(\ref{eq:mag_dipole}) we express the emission rate due to a magnetic dipole as done in nonrelativistic quantum mechanics \cite{1940PhRv...57..225S, 1940ApJ....92..129P}. We follow the result in \citet{1940PhRv...57..225S} for the spontaneous emission coefficient, which shows that transitions are only possible between the same $L$ ($L=L'$) and -- in this case where the spin is $\frac12$ -- we have one of $J$ and $J'$ equal to $L-\frac12$ and the other equal to $L+\frac12$. We write the radial overlap integral explicitly for ease of comparison to Eq.~(\ref{eq:dipole-1}), even though it evaluates to a Kronecker delta. Note that there is an additional factor of $1/4\pi$ due to the conversion from Gaussian units to units where $\epsilon_{0}=1$. Noting there is no radial dependence on the magnetic dipole moment, the emission rate can written as
\begin{eqnarray}
\label{eq:mag_dipole1}
 \frac{d N_\gamma^{(1)}}{dt} &=&  \sum_{J ^{'} M_{J} M_{J}^{'}} \sum_{h h^{'}} \delta_{J^{'},J \pm 1} f_{up,up,h}^{e^{-}} g_{up,up,h^{'}}^{e^{-}}  \frac{1}{3 \pi}(h-h^{'})^{3} \frac{1}{2J+1} \Big{\vert} \bra{J,M_{J}} - \frac{e}{2\mu}(\Vec{L} + 2\Vec{S}) \ket{J^{'}, M^{'}_{J}} \Big{\vert}^{2}
 \nonumber \\
 \ && \times \Bigg{\vert} \int_{0}^{\infty} R^{*}_{Lh}(r)R_{Lh^{'}}(r)dr \Bigg{\vert}^{2}
 \nonumber \\
 \ && = e^{2} \sum_{h h^{'}} f_{up,up,h}^{e^{-}} g_{up,up,h^{'}}^{e^{-}}  \frac{1}{12\pi \mu^{2}}(h-h^{'})^{3} \left\{\begin{array}{lcl}
\frac{L}{2L+1}  & & J^{'} = J-1  \\ 
\frac{L+1}{2L+1} & & J^{'} = J+1  
\end{array} \right\} \Bigg{\vert} \int_{0}^{\infty} R^{*}_{Lh}(r)R_{Lh^{'}}(r)dr \Bigg{\vert}^{2}.
\end{eqnarray}
Note that sum across $M_{J}$ and $M_{J}^{'}$ gives a factor of $L(L+1)/(2L+1)$ which will cancel with the factor of $2J+1 = 2L,2L+2$ in the denominator depending on whether $J=L\pm \frac{1}{2}$, respectively. 

To compare Eq.~(\ref{eq:mag_dipole1}) with Eq.~(\ref{eq:mag_dipole}), we use the normalization given for the $F$'s along with treating the energy levels as a continuous distribution as shown in Eq.~(\ref{eq:dipole_approx}). When we make these replacements we find Eq.~(\ref{eq:mag_dipole1}) and Eq.~(\ref{eq:mag_dipole}) are equivalent.

\section{Charge conjugation in the Schwarzschild spacetime}
\label{app:C}

Charge conjugation is the operation that changes all particles in a given quantum system to their associated antiparticles.
To swap the electron ($\hat b$) operators with positron ($\hat d$) operators, we must take the Hermitian conjugate (since $\psi$ contains $\hat b$ and $\hat d^\dagger$), swap the upper and lower 2 components of the spinor (since the upper components are larger for electrons and the lower for positrons), and swap the spin-up and spin-down sub-components (since we are swapping creation and annihilation operators). We follow the standard approach here (e.g., \S5.2 of \citet{Bjorken1965}), adapted for the Schwarzschild background.

The specific action of the charge conjugation operator $\hat C$ on a fermion field $\psi (x)$ that implements this and has the correct signs to commute with the Hamiltonian is as follows.
We define $\hat C$ to act on the creation and annihilation operators as:
\begin{equation}
\label{eq:charge_conjugation_b_and_d}
    \hat C \hat{b}_{X,k,m,h}\hat C^{-1} = is(-1)^{m+{1}/{2}}\hat{d}_{X,-k,-m,h}
    ~~~{\rm and}~~~
    \hat C\hat{d}_{X,k,m,h}\hat C^{-1} = -is(-1)^{m+{1}/{2}}\hat{b}_{X,-k,-m,h}.
\end{equation}
The complex conjugation rules for the angular modes in Eq.~(\ref{eq:TFGMODE}) are
\begin{equation}
\Theta^{(F)\ast}_{k,m}(\theta,\phi) = 
-is(-1)^{m+1/2}
i\tilde {\boldsymbol\gamma}_{2} {\boldsymbol\beta}\Theta^{(G)}_{-k,-m}(\theta,\phi)
{\rm ~~~~and~~~~}
\Theta^{(G)\ast}_{k,m}(\theta,\phi) = 
-is(-1)^{m+1/2}
i\tilde {\boldsymbol\gamma}_{2} {\boldsymbol\beta}\Theta^{(F)}_{-k,-m}(\theta,\phi),
\end{equation}
where we have used the matrix
\begin{equation}
i\tilde {\boldsymbol\gamma}_{2} {\boldsymbol\beta}
    = \left( \begin{array}{cccc}
  0 & 0 & 0 & -1 \\ 0 & 0 & 1 & 0 \\ 0 & 1 & 0 & 0 \\ -1 & 0 & 0 & 0
    \end{array}\right)
\end{equation}
that implements the aforementioned component swaps.
Then one can show that Eq.~(\ref{eq:fermion_field}) transforms as
\begin{equation}
\label{eq:charge_conjugation}
    \hat C\psi_A(x)\hat C^{-1} = [i\tilde {\boldsymbol\gamma}_{2} {\boldsymbol\beta}]_{AB} \psi^{\dagger}_B (x) .
\end{equation}
Under charge conjugation, we have $k \rightarrow -k$ and $m \rightarrow -m$: this is because we defined the positron operators in the ``Dirac sea'' convention ($\hat d^\dagger_{X,k,m,h}$ fills in a hole in an orbital of $z$-angular momentum $m$, so it creates a positron of $z$-angular momentum $-m$).

Charge conjugation acts as a simple sign flip on the electromagnetic field and mode operators, just as in flat spacetime QED:
\begin{equation}
\hat C A_\mu \hat C^{-1} = -A_\mu
~~~{\rm and}~~~
\hat C \hat a_{X_\gamma,\ell,m_\gamma,\omega,(p)} \hat C^{-1} =
-\hat a_{X_\gamma,\ell,m_\gamma,\omega,(p)}.
\end{equation}
With this sign flip included, it is straightforward to show that $\hat C$ commutes with all three parts of the Hamiltonian ($ H_{\rm Dirac}$, $ H_{\rm \gamma}$, and $ H_{\rm int}$).

Finally, we consider the $I^{+-}$ and $I^{-+}$ terms in Eq.~(\ref{eq: evolve_diss}). The reactions they describe, e.g. $e^{\pm} \rightarrow \gamma~+~e^{\pm}$ are equivalent under charge conjugation, and the fact that $\hat C$ commutes with the interaction Hamiltonian ($\hat C H_{\rm int}\hat C^{-1} = H_{\rm int}$) then implies that $\hat C$ symmetry maps the $I^{+-}$ term in Eq.~(\ref{eq: Q_int}) into the $I^{-+}$ term. Taking into account the factor of $[is(-1)^{m+1/2}]^\ast is'(-1)^{m'+1/2}$ for the mapping of the fermion operators (Eq.~\ref{eq:charge_conjugation_b_and_d}); the $-$ sign from the photon operator; and the $-$ sign from fermion anticommutation (the $\hat b^\dagger$ in the $I^{+-}$ term maps is mapped into the $\hat d^\dagger$, and then it has to be moved to the right of the $\hat d$ to match with the $I^{-+}$ term as written), we must have
\begin{equation}
I^{+-}_{X'k'm',Xkm,X_{\gamma} \ell m_\gamma (p)}(h',h,\omega)
= ss'(-1)^{m-m'}
I^{-+}_{X-k-m,X'-k'-m',X_{\gamma} \ell m_\gamma (p)}(h,h',\omega).
\end{equation}
Using Eq.~(\ref{eq:double-bracket}), the symmetries of the $3j$-symbols, and simplifying using the fact that $\ell$ is an integer whereas $j$, $j'$, $m$, and $m'$ are half-integers, we may write these as a relation among the double-barred coupling coefficients,
\begin{equation}
\label{eq: I_C_symmetry}
\llbracket
I^{+-}_{X'k',Xk,X_{\gamma} \ell (p)}(h',h,\omega)
\rrbracket
= ss'(-1)^\ell
\llbracket
I^{-+}_{X-k,X'-k',X_{\gamma} \ell (p)}(h,h',\omega) \rrbracket
.
\end{equation}

This may also be shown directly from Eq.~(\ref{eq: coupling_int}) by comparing the expressions and swapping the fermion components. To do this, we note that under the swapping $k,h\rightarrow -k',h'$ and $k'h'\rightarrow -k,h$, the radial integrals in Eq.~(\ref{eq: coupling_int}) stay the same; in particular the coefficient $k-k'$ stays the same: $k-k'\rightarrow (-k')-(-k)=k-k'$. For the angular terms, application of the $3j$ symbol symmetry rules to the definitions of $\Delta$ and $\Pi^\pm$ gives
\begin{equation}
\llbracket
\Delta^{-k',-k,\ell}
\rrbracket
= (-1)^{j-j'}
\llbracket
\Delta^{k,k',\ell}
\rrbracket
~~~{\rm and}~~~
\llbracket
\Pi^{-k',-k,\ell\,\pm}
\rrbracket
= ss'(-1)^{\ell}
\llbracket
\Pi^{k,k',\ell\,\pm}
\rrbracket.
\end{equation}
Due to the parity selection rules, $\Delta$ is only non-zero if $ss'(-1)^\ell = (-1)^{j-j'}$. Therefore we conclude that the explicit expressions for the $I$-integrals also obey Eq.~(\ref{eq: I_C_symmetry}).

\bibliography{main}

\end{document}